\documentclass[12pt]{article}

\usepackage[utf8]{inputenc}
\usepackage{cancel}
\usepackage{amssymb}
\usepackage{soul}
\usepackage{tabularx}
\usepackage{xcolor}
\usepackage{booktabs}
\usepackage{subcaption} 
\usepackage{colortbl}
\usepackage{authblk}
\DeclareUnicodeCharacter{00A0}{ }

\newcolumntype{C}{>{\centering\arraybackslash}X}
\usepackage[T1]{fontenc}
\usepackage{epsfig}
\usepackage{latexsym}
\usepackage{graphicx}
\usepackage{amsmath}
\usepackage{amsfonts}   
\usepackage{amssymb}    
\usepackage{float}
\usepackage{bm}
\usepackage{url}
\usepackage{hyperref} 
\usepackage[nodisplayskipstretch]{setspace}
\def\lsim{\raise0.3ex\hbox{$\;<$\kern-0.75em\raise-1.1ex\hbox{$\sim\;$}}}
\def\gsim{\raise0.3ex\hbox{$\;>$\kern-0.75em\raise-1.1ex\hbox{$\sim\;$}}}

\newcommand{\captions}{\sf\caption}
\def    \beq            {\begin{equation}}
\def    \eeq            {\end{equation}}
\def    \bea           {\begin{eqnarray}}
\def    \eea           {\end{eqnarray}}

\def \mn{\mu\nu{\rm SSM}}

\def\g2{{\rm GeV}^2}

\def\sw2{sin^2 \theta_w}

\def\a^tau{\alpha_{\tau}}

\def\beq{\begin{equation}}
\def\eeq{\end{equation}}
\def\beqa{\begin{eqnarray}}
\def\eeqa{\end{eqnarray}}

\newcommand{\newc}{\newcommand}
\newc\BR{BR}
\newc{\akappa}{A_{\kappa} }
\newc\deltagmtwo{\delta (g-2)_{\mu}} 
\newc\deltaamu{\Delta a_{\mu}}

\def\anti{\overline}

\newc{\haa}{BR\(h_1\to a_1 a_1\)}
\newc{\abb}{BR\(a_1\to b\anti{b}\)}
\newc{\hbb}{BR\(h_1\to b\anti{b}\)}
\newc{\abund}{\Omega h^2}
\newc\bsgamma{b\rightarrow s \gamma }
\newc\bxsgamma{\overline{B}\rightarrow X_{s}\gamma}
\newc\brbsgamma{\BR(\overline{B}\rightarrow X_s\gamma)}

\newc{\Fermi}{\textit{Fermi}-}

\allowdisplaybreaks

\usepackage{array, makecell}
\usepackage{boldline}
\usepackage{cite}

\title{\bf{
MeV-GeV~$\gamma$-ray~telescopes~probing axino~LSP/gravitino~NLSP~as dark~matter~in~the~$\mu\nu$SSM 
}}
\author[a]{Germán A. G\'omez-Vargas
\thanks{{\it `Currently, data scientist corporativo at Derco'}}
\thanks{germangomez@derco.cl}}
\author[b,c]{Daniel~E.~L\'opez-Fogliani\thanks{daniel.lopez@df.uba.ar}}
\author[d,e]{Carlos~Mu\~noz\thanks{c.munoz@uam.es}} 
\author[b]{Andres~D.~Perez\thanks{andres.perez@df.uba.ar}}
\affil[a]{Instituto de Astrof\'isica, Pontificia Universidad Cat\'olica de Chile, 
Avenida Vicu\~na Mackenna 4860 Santiago, Chile}
\affil[b]{Instituto de F\'isica de Buenos Aires UBA \& CONICET, Departamento de F\'isica,
Facultad de Ciencia Exactas y Naturales, Universidad de Buenos Aires, 
1428 Buenos Aires, Argentina}
\affil[c]{
{Pontificia Universidad Cat\'olica Argentina, 
1107 Buenos Aires, Argentina}}
\affil[d]{Departamento de F\'{\i}sica Te\'{o}rica, Universidad Aut\'{o}noma de Madrid (UAM),
Campus de Cantoblanco, 28049 Madrid, Spain}
\affil[e]{Instituto de F\'{\i}sica Te\'{o}rica (IFT) UAM-CSIC, 
  Campus de Cantoblanco, 28049 Madrid, Spain}
\date{\today}

\begin{document}
\maketitle
\begin{abstract}

Axino and gravitino are promising candidates to solve the dark matter (DM) problem
in the framework of supersymmetry. In this work, we assume that the axino is the lightest supersymmetric particle (LSP), and therefore contributes to DM. 
%
In the case of $R$-parity violating models, the axino can decay into a neutrino-photon pair with a lifetime much longer than the age of the Universe, yielding a potentially detectable signal. Interestingly, a gravitino next-to-LSP (NLSP) can live enough as to contribute to the relic density. 
%
%
We study both scenarios, only axino LSP as DM, and axino LSP with gravitino NLSP as DM. We carry out the analysis in the context of the $\mu\nu$SSM, which solves the $\mu$ problem and reproduces neutrino data, only adding couplings involving right-handed neutrinos.
In particular, we perform a complete analysis of the relevant parameter space of the model considering constraints from neutrino physics, cosmological observations, and $\gamma$-ray detection. 
We find that the axino or the gravitino can produce a signal detectable by future MeV-GeV $\gamma$-ray telescopes. In addition, in a parameter region where we get a {well-tempered} mixture of both particles, a double-line signal arises as a smoking gun.

\end{abstract}

{\small   Keywords: Supersymmetry, Dark Matter, Gamma Rays} 

\tableofcontents 


\section{Introduction}
\label{sec:intro}

To elucidate the composition of DM is one of the most intriguing enigmas in modern science. The physics and astronomy communities have invested vast efforts in both experimental and theoretical aspects to discover its nature.
Concerning the latter, in $R$-parity conserving (RPC) supersymmetry (SUSY), weakly interacting massive particles (WIMPs)
such as the neutralino~\cite{Goldberg:1983nd,Ellis:1983wd,Krauss:1983ik,Ellis:1983ew} or the right-handed sneutrino (see Refs.~\cite{Cerdeno:2008ep,Cerdeno:2015ega} and references therein), are usual candidates for DM.
However, in $R$-parity violating (RPV) SUSY they
have very short lifetimes, and hence cannot be candidates.  
On the other hand, gravitino ($\psi_{3/2}$) or axino ($\tilde a$) as LSPs 
can be valid superWIMP DM candidates. 
Although they also decay as neutralinos or sneutrinos, their lifetimes turn out to be much longer than the age of the Universe.
In the case of the gravitino, its lifetime is suppressed both by the gravitational interaction and by the small RPV 
couplings~\cite{Borgani:1996ag,Takayama:2000uz}, whereas for the axino in addition to the latter it is also suppressed by the Peccei-Quinn (PQ) scale~\cite{Covi:2009pq}.
Besides, gravitino or axino decays produce $\gamma$ rays which could be observed in $\gamma$-ray telescopes. This was analyzed for the 
gravitino in Refs.~\cite{Takayama:2000uz,Buchmuller:2007ui,Bertone:2007aw,Ibarra:2007wg,Ishiwata:2008cu,Choi:2010xn,Choi:2010jt,Diaz:2011pc,Restrepo:2011rj,Kolda:2014ppa,Bomark:2014yja} 
in the context of bilinear/trilinear RPV models~\cite{Barbier:2004ez},
and in Refs.~\cite{Choi:2009ng,GomezVargas:2011ph,Albert:2014hwa,GomezVargas:2017} in the
`$\mu$ from $\nu$' supersymmetric standard 
model ($\mu\nu$SSM)~\cite{propuvSSM}.
Similar analyses for the axino in bilinear/trilinear RPV models were carried out 
in 
Refs.~\cite{Kim:2001sh,Hooper:2004qf,Chun:2006ss,Endo:2013si,Kong:2014gea,Choi:2014tva,Liew:2014gia,Colucci:2015rsa,Bae:2017tqn,Colucci:2018yaq}.

On the other hand, in Ref.~\cite{Hamaguchi:2017}, the authors studied the cosmology of an RPC example of 
decaying dark matter (DDM) 
scenarios~\cite{Olive:1984bi,Chun:1993vz,Kim:1994ub,Asaka:2000ew,Ichikawa:2007jv,Hasenkamp:2011em,Graf:2013xpe}, which could relax 
potential tensions
between the standard $\Lambda$CDM model and cosmological observations~\cite{Berezhiani:2015,Chudaykin:2016,Poulin:2016,Chudaykin:2017,Bringmann:2018}, considering 
both  
$\psi_{3/2}$ and $\tilde a$ as DM candidates.
In that case,
their masses $m_{3/2}$ and $m_{\tilde a}$, respectively, are model dependent and 
can be of the same order or several orders of magnitude different
in realistic 
scenarios~\cite{Kawasaki:2013ae,Goto:1991gq,Chun:1992zk,Chun:1995hc,Kim:2012bb} such as in supergravity. Therefore, if the axino (gravitino) is the LSP the gravitino (axino) can become the NLSP. As a consequence, the NLSP decays into the LSP plus an axion.

In the present work, 
we will consider axino LSP as DM in the context of RPV, for two particularly interesting scenarios. First, only axino LSP as DM. This can be achieved if the NLSP is not the gravitino, but another SUSY particle which has a short lifetime, as happens in the case of RPV models.
Second, a DDM scenario with axino LSP and gravitino NLSP.
In both scenarios, we will study their cosmological properties as well as 
associated $\gamma$-ray constraints on spectral lines coming from current detectors, and prospects for future $\gamma$-ray space missions.
Concerning the RPV model, we will concentrate on the $\mu\nu$SSM~\cite{propuvSSM,analisisparam} given its phenomenological interest. 

The $\mn$ introduces new couplings in the superpotential with respect to 
the RPC minimal supersymmetric standard model (MSSM)~\cite{Nilles:1983ge,Haber:1984rc,Martin:1997ns} and Next-to-MSSM (NMSSM)~\cite{Ellwanger:2009dp}. These couplings
involve right-handed neutrino superfields to solve 
the $\mu$-problem of SUSY,
while simultaneously are able to reproduce at tree level the observed neutrino masses
and mixing angles~\cite{propuvSSM,analisisparam,Ghosh:2008yh,neutrinocp,Ghosh:2010zi}.
The latter is obtained through a generalized electroweak-scale seesaw mixing left- and right-handed neutrinos with neutralinos.
In addition, 
the extrapolation of the usual stringent bounds on sparticle masses in RPC SUSY to the $\mn$ is not applicable. 
For example, it was shown in Refs.~\cite{Lara:2018rwv,Kpatcha:2019gmq} that the LEP lower bound on masses of slepton LSPs of about 90 GeV obtained in 
trilinear RPV~\cite{Abreu:1999qz,Abreu:2000pi,Achard:2001ek,Heister:2002jc,Abbiendi:2003rn,Abdallah:2003xc} is not valid in the $\mn$.
For the bino LSP,\footnote{The phenomenology of a neutralino LSP was analyzed 
in the past in 
Refs.~\cite{Ghosh:2008yh,Bartl:2009an,Ghosh:2012pq,Ghosh:2014ida}.}
only a small region of the parameter space of the $\mn$ 
was excluded~\cite{Lara:2018zvf} when the left sneutrino is the NLSP and hence a suitable source of binos. In particular, this was the case of
the region of bino (sneutrino) masses $110-150$ ($110-160$) GeV.
It is worth pointing out here that gravitino or axion DM do not alter these collider signals, since effectively any NLSP such as sneutrino, bino, etc., behaves like an usual LSP in RPV models decaying fast through RPV channels~\cite{Fidalgo:2011ky}.
The Higgs sector of the $\mn$ is also very interesting 
phenomenologically~\cite{analisisparam,Ghosh:2014ida,Biekotter:2017xmf,Biekotter:2019gtq,Kpatcha:2019qsz}, 
since there is a substantial mixing among
the three right-handed sneutrinos and the doublet-like Higgses.
Cosmological issues in the model have also been considered, and in particular the
generation of the baryon asymmetry of the Universe was studied in detail in 
Ref.~\cite{Chung:2010cd}, with the interesting result that electroweak baryogenesis can be realized.


In Refs.~\cite{Choi:2009ng,GomezVargas:2011ph,Albert:2014hwa,GomezVargas:2017}, it was shown
that the mixing mentioned above between neutralinos, in particular the photino and left-handed neutrinos in the neutral fermion mass matrix, has crucial consequences on the gravitino DM phenomenology of the $\mn$, and as we will show in this work, also in the axino DM phenomenology. 
In particular, axino and gravitino can decay to a photon and a neutrino through RPV terms producing a mono-energetic $\gamma$-ray signal. The energy ranges of \Fermi LAT and previous missions COMPTEL and EGRET, lie in the ballpark for these candidates, and can therefore test the axino/gravitino DM hypothesis. Besides, planned detectors aimed to explore $\gamma$ rays, such as e-ASTROGAM~\cite{eAstrogamAngelis:2017} and AMEGO~\cite{AmegoCaputo:2017}, will feature a 2-3 order of magnitude increase in sensitivity and an improvement in the energy resolution of the $\gamma$-ray sky in comparison to COMPTEL and EGRET in the MeV to GeV range. 

We will analyze the allowed parameter space considering the two scenarios mentioned before: only axino LSP as DM; and axino LSP with gravitino NLSP as DM, with the latter decaying to the former (plus an axion). We will find that gravitino and axino masses, as well as the PQ scale, play a crucial role in defining the characteristics of the model. 
In addition, 
the photino-neutrino RPV parameter has to be considered imposing the constraints from neutrino physics. Finally, as a benchmark for future $\gamma$-ray missions we will consider the performance of e-ASTROGAM, and we will show that such kind of instruments can probe a significant portion of the parameter space. Besides, if axino and gravitino can coexist, each one can give rise to a spectral line detectable by e-ASTROGAM, producing a `smoking gun' signal in the form of two $\gamma$-ray lines that are difficult to mimic with standard astrophysical processes.


The paper is organized as follows. 
In Section~\ref{sec:axinodm}, we will discuss the scenario with only axino LSP as DM.
We will show the decay rate of axino DM into photon plus neutrino, as well as the amount of the associated relic density. Then, we will analyze the $\gamma$-ray flux produced in this scenario, showing the exclusion limits and prospects for detection.
In Section~\ref{sec:DM Scenario}, we will discuss the DDM scenario including axinos and gravitinos. We will show the gravitino NLSP decay rates into photon plus neutrino and into axino LSP plus axion, and its contribution to the relic density.
%
Armed with these results, we will be able to fully explore 
our multicomponent DM scenario along with its parameter space allowed by cosmological observations. 
Finally, as in the previous section, we will analyze the $\gamma$-ray flux of this scenario, exclusion limits and prospects for detection. 
The conclusions are left for Section~\ref{sec:conclusions}.


\section{Axino LSP as dark matter
}
\label{sec:axinodm}

In the framework of supergravity, the
axino has an interaction term in the Lagrangian with photon and photino. As discussed in the
introduction, in the presence of RPV photino and left-handed neutrinos are mixed in the neutral fermion mass matrix, and therefore the axino LSP is able to decay through this interaction term into photon and neutrino.  
This has relevant implications because the $\gamma$-ray signal is a sharp line with an energy ${m_{\tilde{a}}}/{2}$, that could be detected in $\gamma$-ray space telescopes 
such as \Fermi LAT, or in future MeV-GeV telescopes such as the proposed e-ASTROGAM.

\subsection{Axino decay}
\label{axinodecay}

Axino decay width into photon-neutrino through RPV couplings is given by~\cite{Covi:2009pq}:
\bea
\Gamma(\tilde{a}\rightarrow
\gamma
\nu_i)
\simeq\frac{m_{\tilde{a}}^3}{128\pi^3 f_a^2}\alpha_{em}^2C_{a\gamma\gamma}^2|U_{\tilde{\gamma} \nu}|^2,
\label{decay2bodyaxino}
\eea
where 
$\Gamma(\tilde{a}\rightarrow \gamma \nu_i)$ denotes a sum of the partial decay widths into $\nu_i$ and $\bar\nu_i$,
$C_{a\gamma\gamma}$ is a model dependent constant of order 
unity, $\alpha_{em}=e^2/4\pi$, $f_a$ is the PQ scale, 
and the mixing parameter $|U_{\tilde{\gamma} \nu}|$ determines the photino content of the neutrino 
\bea
\left|U_{\tilde{\gamma} \nu}\right|^2= \sum^3_{i=1}\left|N_{i1} \, \cos\theta_W +  N_{i2} \, \sin\theta_W\right|^2.
\label{photino}
\eea
Here $N_{i1} (N_{i2})$ is the bino (wino) component of the $i$-$th$ neutrino, and $\theta_{W}$ is the weak mixing angle. 
As obtained in Refs.~\cite{Choi:2009ng,GomezVargas:2017}, performing scans in the low-energy parameters of the $\mn$ in order to reproduce the observed neutrino masses and mixing angles, natural values of $|U_{\tilde{\gamma} \nu}|$ are in the range 
\begin{equation}
10^{-8} \lesssim |U_{\widetilde{\gamma}\nu}| \lesssim 10^{-6}.
\label{relaxing}
\end{equation}
Relaxing some of the assumptions such as 
an approximate GUT relation for gaugino masses and/or TeV scales, the lower bound can be smaller:
\begin{equation}
10^{-10} \lesssim |U_{\widetilde{\gamma}\nu}| \lesssim 10^{-6}.
\label{relaxingmore}
\end{equation}

As we can see in Eq.~(\ref{decay2bodyaxino}), the axino decay is suppressed both, by the small RPV mixing parameter $|U_{\tilde{\gamma} \nu}|$, and by the large PQ scale $f_a \gsim 10^{9}$ GeV as obtained from the observation of SN1987A~\cite{Kawasaki:2013ae}.
This gives rise to a lifetime longer than the age of the Universe,
${\tau}_{\tilde{a}}\gg t_{today}\sim 10^{17}$ s, {with 
\begin{equation}
{\tau}_{\tilde{a}} = \Gamma^{-1}(\tilde{a}\rightarrow
\gamma
\nu_i) 
\simeq
3.8\times 10^{28}\, {s}
\left(\frac{f_a}{10^{13}\, \mathrm{GeV}}\right)^2
\left(\frac{10^{-8}}{|U_{\widetilde{\gamma}\nu}|}\right)^2
\left(\frac{0.1\, \mathrm{GeV}}{m_{\tilde{a}}}\right)^{3},
\label{axinolifetime}
\end{equation}
where in the last equality we have assumed $C_{a\gamma\gamma}=1$.} 





\subsection{Axino relic density}
\label{axinorelic}

Although axino decays, we have shown in the previous subsection that $\tau_{\tilde a}\gg t_{today}$, and therefore in a very good approximation we can consider that its relic density coincides with the would-be axino relic density if it were stable and would not undergo through the decay process.
For axinos, this relic density depends heavily on the axion model considered. 
%
%
In the framework of the KSVZ model~\cite{Kim:1979,Shiftman:1980}, the axino production is dominated by the scattering of gluons and gluinos and its relic density from thermal production turns out to 
be~\cite{Brandenburg:2004,Strumia:2010}
%
\begin{equation}
\Omega^{\text{TP}}_{\tilde{a}}h^2\simeq 0.3 \ (g_3 (T_R))^4 \ \left(\frac{F(g_3(T_R))}{23}\right) \left( \frac{m_{\tilde{a}}}{1 \text{ GeV}} \right) \left(\frac{T_R}{10^4 \text{ GeV}}\right) \left(\frac{10^{12}\text{ GeV}}{f_a}\right)^2,
\label{relicaxinos}
\end{equation}
where $T_R$ is the reheating temperature after inflation, $g_3 $ is the running $SU(3)$ coupling, and the rate function $F(g_3(T_R))$ describes the axino production rate with $F\simeq 24-21.5$ for $T_R\simeq 10^4-10^6$~GeV~\cite{Strumia:2010}. 
For our numerical computation we will use 
$F\simeq 23$. Other values will not change significantly the final results.

To continue we must address the axion production. This comes from the misalignment mechanism, and 
therefore the axion cold DM relic density can be accounted by
\begin{equation}
\Omega_{a}h^2\simeq 0.18\ \theta^2_i \left(\frac{f_a}{10^{12}\text{~GeV}}\right)^{1.19},
\label{relicaxions}
\end{equation}
where $\theta_i$ is the initial misalignment angle. Since we are interested in studying the scenario with axino as the only component of the DM, we can
always set the axion primordial relic negligible choosing an appropriate value for $\theta_i$ if needed, i.e. when $f_a\gsim 10^{12}$~GeV.
Nevertheless, it would be convenient to work with the upper bound $f_a\leq 10^{13}$ GeV to avoid too much tuning.



Considering therefore that the axino is the only component of DM, the relic density given by Eq.~(\ref{relicaxinos}) 
is proportional to the reheating temperature.
Clearly, given an axino mass and PQ scale, adjusting $T_R$
one can get the measured value of the relic density by the Planck Collaboration~\cite{Aghanim:2018eyx},
$\Omega_{cdm}^{\text{Planck}}h^2\simeq 0.12$.
In particular, 
one obtains
\begin{equation}
T_R
\simeq \frac{0.4}{(g_3 (T_R))^4}\times {10^4\ \text{GeV}}\ \left(\frac{1\ \text{GeV}}{m_{\tilde{a}}}\right) 
\left(\frac{f_a}{10^{12}\ \text{GeV}}\right)^2.
\label{reheating}
\end{equation}
For example, for $m_{\tilde{a}}=0.1$ GeV and $f_a=10^{12}$ GeV one needs $T_R\simeq 7.2 \times 10^4$ GeV.
Also, assuming the conservative limit $T_R \gtrsim 10^4$ GeV, an upper bound for $m_{\tilde{a}}$ is obtained from Eq.~(\ref{reheating}) for each value of $f_a$:
\begin{equation}
m_{\tilde{a}}\lsim 
0.5\ \text{GeV}\ \left(\frac{f_a}{10^{12}\ \text{GeV}}\right)^2.
\label{reheating2}
\end{equation}
%
For example, for $f_a=10^{13}, 10^{12}, 10^{11}$ GeV one obtains the upper bounds  $m_{\tilde{a}}\lsim 50, 0.5, 0.005$ GeV, respectively.
%
%
Note that to use the lower bound $f_a\geq 10^{11}$ GeV is convenient to avoid a too small axino mass, beyond the reach of proposed detectors (see Fig.~\ref{figaxinoalone1} below). Thus, throughout this work we will adopt the following range for the PQ scale:
\begin{equation}
10^{11} \leq f_a \leq 10^{13}\ \text{GeV}.
\label{pqscale}
\end{equation}

On the other hand, in the case of the DFSZ axion model~\cite{Dine:1981,Zhitnitsky:1980} the axino production is dominated by axino-Higgs-Higgsino and/or axino-quark-squark interactions.  For reheating temperatures above $10^4$~GeV
as the ones used in this work, the axino relic density is in a good approximation independent of the reheating temperature, and it turns out to be~\cite{Bae:2011}
\begin{equation}
\Omega^{\text{TP}}_{\tilde{a}}h^2\simeq 20.39 \left( \frac{m_{\tilde{a}}}{1 \text{ GeV}} \right) \left(\frac{10^{12}\text{ GeV}}{f_a}\right)^2.
\label{relicaxinosDFSZ}
\end{equation}
Thus, if the axino is the only component of DM a fixed mass is obtained in this case for a given $f_a$:
\begin{equation}
m_{\tilde{a}}\simeq 6\ \text{MeV}\ \left(\frac{f_a}{10^{12}\ \text{GeV}}\right)^2.
\label{maDFSZ}
\end{equation}
For example, for $f_a=10^{13}, 10^{12}, 10^{11}$ GeV one obtains $m_{\tilde{a}}\simeq 600, 6, 0.06$ MeV, respectively.

In Subsec.~\ref{results}, we will discuss the prospects for detection 
of monochromatic lines coming from axino decay. We will see that the 
DFSZ model covers a subset of the relevant parameters with respect to the 
KSVZ model.
Hence, in the next section we will continue working with the KSVZ model 
for the sake of generality.

\subsection{$\gamma$-ray flux from axino decay}
\label{gammarayaxino}

The constraints set by detectors such as \Fermi LAT to the $\gamma$-ray emission from DM, 
place lower limits to the axino lifetime as the source of $\gamma$-ray radiation. 
The differential flux of $\gamma$ rays from DM decay in the Galactic halo is calculated by integrating its distribution around us along the line of sight: 
\begin{equation}
 \frac{d\Phi_{\gamma}^{\text{halo}}}{dEd\Omega}=\frac{1}{4\,\pi\,\tau_{\textit{DM}}\,m_{\textit{DM}}}\,\frac{1}{\Delta\Omega}\,\frac{dN^{\text{total}}_{\gamma}}{dE} D_{dec}\ ,
 \label{eq:decayFlux}
\end{equation}
where $\tau_{\textit{DM}}$, $m_{\textit{DM}}$ are the lifetime and mass of the DM particle respectively, $\frac{dN^{\text{total}}_{\gamma}}{dE}$ is the total number of photons produced in DM decay, $\Delta \Omega$ is the solid angle supported by the region of interest (ROI), i.e. the region of the sky that we are studying, and $D_{dec}$ is the so-called $D$-$factor$, involving astrophysical parameters. The latter is defined as
\begin{equation}
 D_{dec}=\int_{\Delta\Omega}\!\!\cos
 b\,db\,d\ell\int_0^{\infty}\!\! ds\,\rho_{\text{halo}}(r(s,\,b,\,\ell))\ ,
 \label{eq:Jfactor}
\end{equation}
where $b$ and $\ell$ denote the Galactic latitude and longitude,
respectively, and $s$ denotes the distance from the Solar System. The radius $r$ in the DM halo density profile of the Milky Way, $\rho_{\text{halo}}$, is expressed in terms of these Galactic coordinates.

It is straightforward to apply now this result to the case of axino LSP as DM,
using the formulas of Sec.~\ref{axinodecay}.

\begin{figure}[t!]
\begin{center}
 \begin{tabular}{cc}
 \hspace*{-4mm}
 \epsfig{file=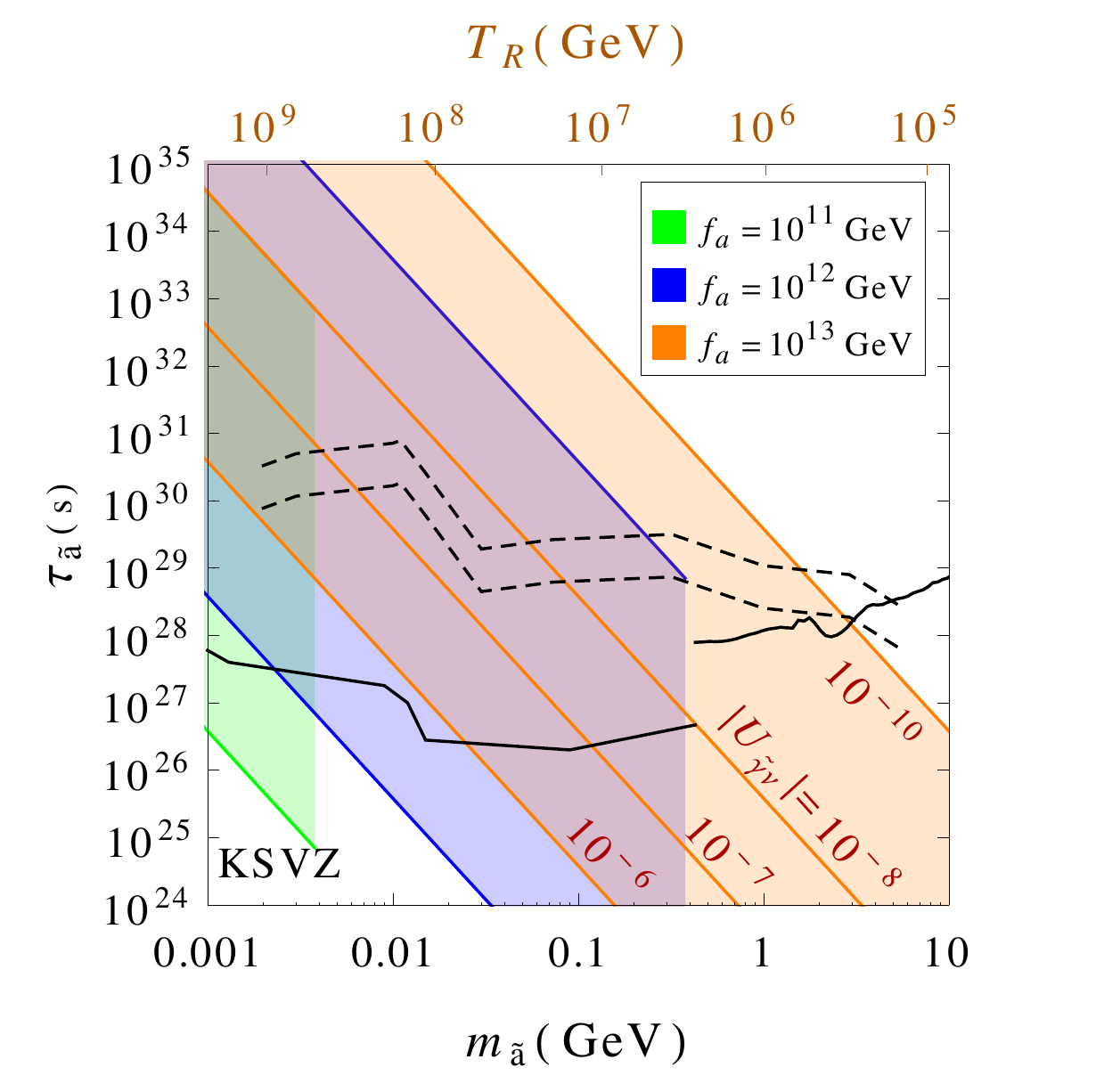,height=8cm} 
       \epsfig{file=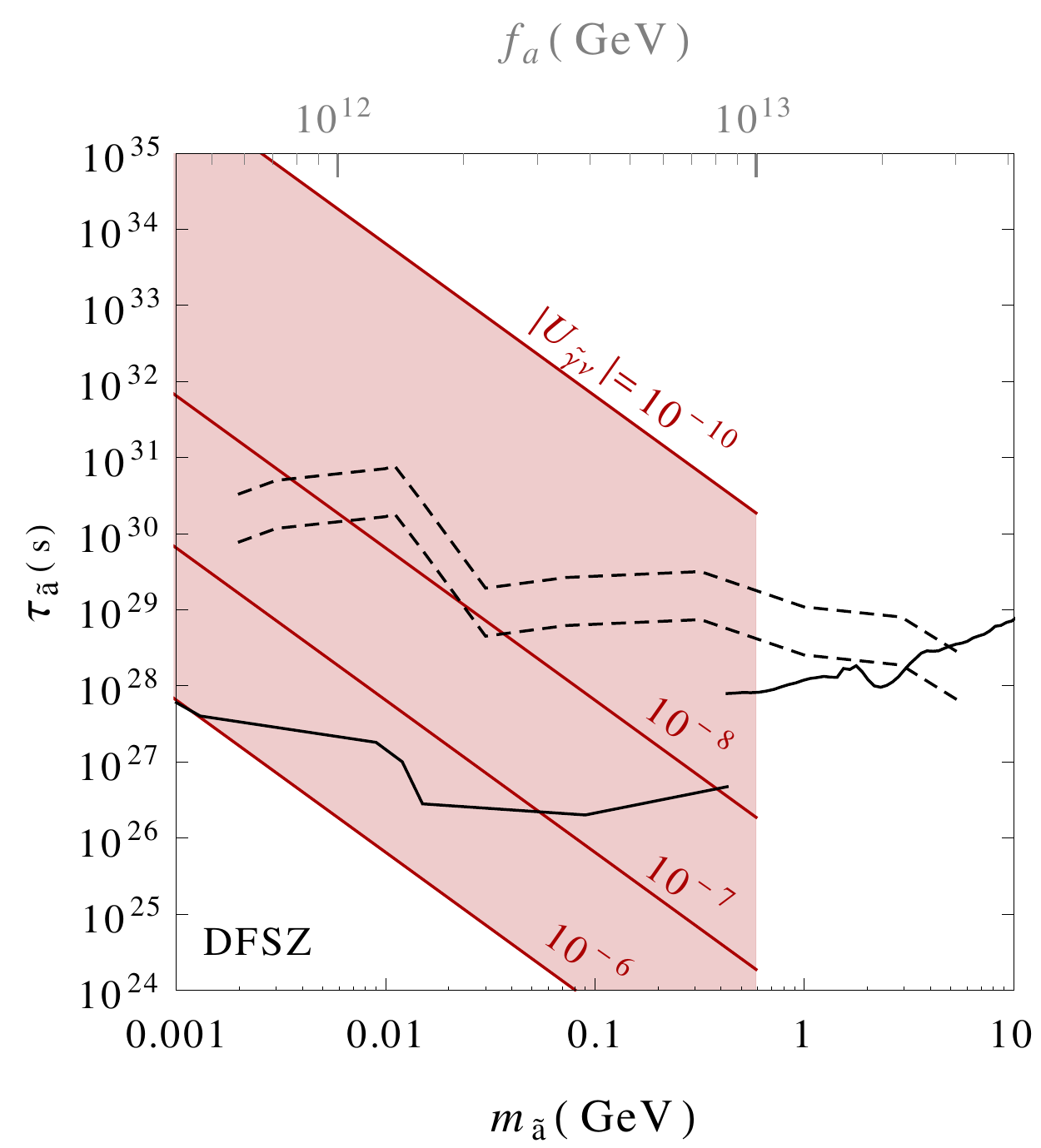,height=7.8cm}   
    \end{tabular}
    \captions{Constraints on lifetime versus mass for axino DM. 
The region below the black solid line on the left (right) is excluded by line searches in the Galactic halo by COMPTEL (\Fermi LAT~\cite{Ackermann:2015lka}). The region below the upper (lower) black dashed line could be probed by e-ASTROGAM~\cite{eAstrogamAngelis:2017} with observations of the Galactic center assuming Einasto B (Burkert) DM profile. Left panel (KSVZ axion model):
In the orange region,
orange solid lines correspond to the predictions of the $\mn$ for several representative values of 
$|U_{\tilde{\gamma} \nu}|$, for the case $f_a=10^{13}$ GeV.
For the cases $f_a=10^{12}$ and $10^{11}$ GeV, associated to blue and green regions, respectively, only the lines corresponding to the lower and upper limits from neutrino physics, $10^{-10} \leq|U_{\tilde{\gamma} \nu}|\leq 10^{-6}$, are shown. 
The reheating temperature versus axino mass shown in orange in the upper part of the figure, corresponds to the case $f_a=10^{13}$ GeV. For other values of $f_a$, $T_R$ can be 
straightforwardly obtained from Eq.~(\ref{reheating}). 
The upper bound on $m_{\tilde{a}}$ for each region is obtained from Eq.~(\ref{reheating2}). Right panel (DFSZ axion model): In the upper part of the figure, the PQ scale versus axino mass using Eq.~(\ref{maDFSZ}) is shown in gray. The red region corresponds to the predicted $\mn$ parameter space for the same representative values used in the left panel. 
}
    \label{figaxinoalone1}
\end{center}
\end{figure}

\subsection{
Results
}
\label{results}



Since in RPV models the axino decays producing a monochromatic
photon with an energy ${m_{\tilde{a}}}/{2}$, one can constrain their parameter space 
with $\gamma$-ray observations. 
Actually, there are model independent constraints on DM decays.
In Fig.~\ref{figaxinoalone1},
the regions below the black solid lines are excluded by line searches 
by COMPTEL and \Fermi LAT~\cite{Ackermann:2015lka}. The black dashed lines correspond to the projected e-ASTROGAM sensitivity~\cite{eAstrogamAngelis:2017}, where we have considered the following DM profiles for the observations of
a region of interest of 10$^\text{o} \times$10$^\text{o}$ around the Galactic center: NFW, Moore, Einasto, Einasto B and Burkert. In particular, Einasto B (Burkert) is the most (least) stringent and corresponds in 
the figure 
to the upper (lower) dashed line.

Using the results from previous subsections, we also show
for the KSVZ axion model in the left panel of Fig.~\ref{figaxinoalone1} with orange solid lines, the values of the parameters predicted by the $\mn$ using Eq.~(\ref{axinolifetime}) with $f_a=10^{13}$ GeV for several representative values of $|U_{\tilde{\gamma} \nu}|$.
For the cases $f_a~=~10^{11}$ and $10^{12}$ GeV we show only the lines corresponding to the lower and upper limits from neutrino physics of Eq.~(\ref{relaxingmore}).
As we can see, values of the axino mass larger than 3 GeV are already disfavored by 
\Fermi LAT. In addition, a significant region of the parameter space of axino DM lies
in the ballpark of future $\gamma$-ray missions such as the proposed e-ASTROGAM, allowing to explore masses and lifetimes in the ranges $2$ MeV$-3$ GeV 
and $2 \times 10^{26}-8 \times 10^{30}$ s, respectively.

Let us finally remark that the upper bound on the axino mass 
for each value of 
$f_{\tilde a}$, e.g. $m_{\tilde{a}}\lsim 0.005, 0.5$ GeV for $f_a=10^{11}, 10^{12}$ GeV, respectively, is obtained 
from Eq.~(\ref{reheating2}) under the conservative limit $T_R \gsim 10^4$ GeV.

On the other hand, the red region in the right panel of Fig.~\ref{figaxinoalone1} corresponds to the predictions of the $\mn$ considering a DFSZ axion model. Unlike the KSVZ model, the axino relic density is 
independent on $T_R$, as already discussed in Eq.~(\ref{relicaxinosDFSZ}). This allows us to simplify the figure.
In the upper part of it, the PQ scale versus axino mass using Eq.~(\ref{maDFSZ}) is shown in gray.

It is worth noticing here that
the allowed red region in the right panel, obtained with a DFSZ model, is in fact included in the allowed region of the left panel for a KSVZ model.
For each value of $m_{\tilde{a}}$ given below Eq.~(\ref{maDFSZ}), we can identify the allowed range of $\tau_{\tilde{a}}$ in the corresponding $f_a$ colored region
of the left panel.
We can also extrapolate the results for intermediate values of $f_a$. Therefore, overlaying both panels we can see that the DFSZ allowed region represents a subset of the KSVZ region, and hence is more restrictive. This is obviously expected, since the former model has one degree of freedom less, $T_R$, in order to obtain the correct relic density. Thus, in the rest of this work we will focus on the KSVZ model to explore axino DM with a broad approach.

\section{Axino LPS and gravitino NLSP as dark matter
}
\label{sec:DM Scenario}

As discussed for the axino,
the gravitino has also an interaction term in the Lagrangian with photon and photino, and therefore in the presence of RPV it is able to decay into photon and neutrino producing
a sharp $\gamma$-ray line with an energy ${m_{3/2}}/{2}$.
In addition, the gravitino NLSP can decay to axino LSP and axion.
We will study the implications of this scenario for DM and its detectability.

Concerning the gravitino mass, let us point out that in supergravity models it is related to the mechanism of SUSY breaking. In particular, in 
gravity-mediated SUSY breaking models, where the soft scalar masses are typically determined by the gravitino mass, it is sensible to expect the latter in the range GeV-TeV~\cite{Brignole:1997dp}, i.e. around the electroweak scale. However, specific Kahler potentials and/or superpotentials of the supergravity theory could allow for different situations, producing gravitinos with masses several orders of magnitude smaller than the electroweak scale. This is e.g. the case of no-scale supergravity models, where the gravitino mass is decoupled from the rest of the SUSY particle spectrum, and hence is possible to assign for it a mass much smaller than the electroweak 
scale~\cite{Ellis:1984kd}. On the other hand, very small gravitino masses with respect to the electroweak scale are obtained in gauge-mediated SUSY breaking models~\cite{Giudice:1998bp}. Also, e.g. in F-theory GUTs with the latter SUSY breaking mechanism working, one can obtain a gravitino mass 
of about $10-100$ MeV~\cite{Heckman:2009mn}.
Given the model dependence of the gravitino mass, we consider appropriate for our phenomenological work below not to choose a specific underlying supergravity model, and treat the gravitino mass as a free parameter.

\subsection{Gravitino NLSP decays
}
\label{subsec:Gravitino decay}

Gravitino partial decay width into photon-neutrino through RPV couplings is given by~\cite{Borgani:1996ag,Takayama:2000uz}:
\bea
\Gamma(\psi_{3/2}\rightarrow
\gamma
\nu_i)
\simeq\frac{m_{3/2}^3}{32\pi M_{P}^2}|U_{\tilde{\gamma} \nu}|^2,
\label{decay2bodygravitino}
\eea
where 
$\Gamma(\psi_{3/2}\rightarrow \gamma \nu_i)$ denotes a sum of the partial decay widths into $\nu_i$ and $\bar\nu_i$, and
$M_P\approx 2.43 \times 10^{18}$~GeV is the reduced Planck mass.
This
is the dominant decay for a gravitino LSP in the context of the $\mu\nu$SSM, 
and 
is suppressed both by the small RPV mixing parameter and by the scale of the gravitational interaction.
{We can compare Eq.~\ref{axinolifetime} with this equation written as
\begin{equation}
{\Gamma^{-1}}
(\psi_{3/2}\rightarrow
\gamma
\nu_i)
\simeq 3.8\times 10^{30}\, {s}
\left(\frac{10^{-8}}{|U_{\widetilde{\gamma}\nu}|}\right)^2
\left(\frac{1\, \mathrm{GeV}}{m_{3/2}}\right)^{3}\ ,
\label{gravitinolifetime_LSP}
\end{equation}
to obtain the following relation between decay widths:
\begin{equation}
\frac{\Gamma(\tilde{a}\rightarrow\gamma\nu_i)}
{{\Gamma}(\psi_{3/2}\rightarrow\gamma\nu_i)}
\simeq 10^{5}\,
\left(\frac{10^{13}\, \mathrm{GeV}}{f_a}\right)^2
r_{\tilde a}^3\ ,
\label{relation}
\end{equation}
where
\bea 
r_{\tilde{a}}\equiv \frac{m_{\tilde{a}}}{{m_{3/2}}}.
\label{ra}
\eea
As we can see, 
$\Gamma(\tilde{a}\rightarrow\gamma\nu_i)$ is typically larger than 
$\Gamma(\psi_{3/2}\rightarrow\gamma\nu_i)$ unless $r_{\tilde a}$ is very small.
In particular, for $f_a=10^{13}, 10^{12}, 10^{11}$ it has to be smaller than about 0.02, 0.004, 0.001, respectively. This result will be useful for our discussion in 
Subsect.~\ref{subsec: Prospects}.}

{Since in the framework of supergravity the gravitino has an interaction term with axino and axion, we have also to consider the RPC partial decay width~\cite{Hamaguchi:2017}
}
\begin{equation}
\Gamma(\psi_{3/2} \rightarrow \tilde{a} \, a) \simeq \frac{m_{3/2}^3}{192\pi M_P^2}(1-r_{\tilde{a}})^2
(1-r_{\tilde{a}}^2)^3, 
\label{decaytoaa}
\end{equation}
where the axion mass has been neglected.
Clearly, it dominates over the one in Eq.~(\ref{decay2bodygravitino}), and therefore the gravitino
lifetime can be approximated as 
\begin{equation}
{\tau}_{3/2}\simeq {\Gamma}^{-1}
(\psi_{3/2} \rightarrow a \, \tilde{a})\simeq  2.3 \times 10^{15} s \: \left(\frac{1\, \text{GeV}}{m_{3/2}} \right)^3,
\label{gravitinolifetime1}
\end{equation}
where to write the last formula we have 
neglected the contribution of $r_{\tilde{a}}$ in Eq.~(\ref{decaytoaa}) which is
valid when
$m_{\tilde a}\ll m_{3/2}$.



%
%

At this point it is important to notice that although
${\Gamma^{-1}}(\psi_{3/2}\rightarrow\gamma\nu_i) \gg t_{today}$, this 
does not hold for ${\tau}_{3/2}$, {implying that the equations for the relic density that will be obtained below are affected by this result.}

\subsection{Axino and gravitino relic density
}
\label{subsec: relic}

To compute axino relic density we need to consider thermal and non-thermal production mechanisms. The latter, in our multicomponent scenario, is related to the decay of the gravitino NLSP {involving its number density and lifetime}. This axino production would not undergo re-annihilation since the Planck mass suppresses gravitino or axino interactions. 

Unlike the axino case in Subsec.~\ref{axinorelic}, whose lifetime is longer than the age of the Universe, and therefore its relic density can be approximated as that from thermal production, the gravitino has a smaller lifetime and one has to consider that its density changes in time with the result
\bea
\Omega_{3/2}h^2 &=& \Omega_{3/2}^{\text{TP}}h^2 e^{-(t_{\text{today}}-t_0)/ \tau_{3/2}},
\label{NLSPrelicgeneral0}
\eea
where $t_0$ is the time when the gravitinos are thermally produced,
and
$\Omega_{3/2}^{\text{TP}}h^2$ corresponds to 
the would-be gravitino NLSP relic density if it were stable and would not undergo through the decay process. The latter is given by~\cite{Bolz:2000fu,Pradler:2006qh,Rychkov:2007uq}
\begin{equation}
\Omega^{\text{TP}}_{3/2}h^2\simeq 0.02\left(\frac{T_R}{10^5 \text{ GeV}}\right)\left(\frac{1 \text{ GeV}}{m_{3/2}}\right)\left(\frac{M_3(T_R)}{3\text{ TeV}}\right)^2\left(\frac{ \gamma(T_R) / (T_R^6/M_P^2)}{0.4}\right).
\label{relicgravitinos}
\end{equation}
Here,
$M_3(T_R)$ is the running gluino mass, and the last factor parametrizes the effective production rate ranging $\gamma(T_R) / (T_R^6/M_P^2)\simeq 0.4-0.35$ for $T_R\simeq 10^4-10^6$~GeV~\cite{Rychkov:2007uq}.
For our numerical computation we will use $M_3(T_R) \simeq 3$ TeV and $\gamma(T_R)/(T_R^6/M_P^2) \simeq 0.4$. Other values will not modify significantly our results.
Assuming as in the previous section  the conservative limit $T_R \gtrsim 10^4$ GeV, a lower limit for the gravitino mass from the measured value of the relic density is obtained, $m_{3/2} \gtrsim 0.017 \text{ GeV}$.


Taking the above into account, the density for axino LSP is now 
\bea
\Omega_{\tilde{a}}h^2 &=& \Omega_{\tilde{a}}^{\text{TP}}h^2  + \Omega_{\tilde{a}}^{\text{NTP}}h^2,
\label{LSPrelicgeneral}
\eea
where $\Omega_{\tilde{a}}^{\text{TP}}h^2$ is given in Eq.~(\ref{relicaxinos}), and the term
$\Omega_{\tilde{a}}^{\text{NTP}}h^2$ takes into account the non-thermal production via gravitino decay:
\bea
\Omega_{\tilde{a}}^{\text{NTP}}h^2  
= 
r_{\tilde a}\
\Omega_{3/2}^{\text{TP}}h^2\ \left(1-e^{-(t_{\text{today}}-t_0)/ \tau_{3/2}}\right).
\label{LSPrelicgeneral2}
\eea
It is worth noticing that the factor $r_{\tilde a}$
takes into account whether the LSP non-thermally produced is either relativistic or non-relativistic, as we are only interested~in~cold~DM.
Obviously, if $\tau_{3/2}\ll t_{today}$, we get the usual relations~\cite{Covi:1999ty,Choi:2011yf,Roszkowski:2014}:
\bea
\Omega_{3/2}h^2 &\approx& 0,\\
\Omega_{\tilde{a}} h^2 \; &\approx& \Omega_{\tilde{a}}^{\text{TP}}h^2 + 
r_{\tilde a}\
\Omega_{3/2}^{\text{TP}}h^2.
\label{usual}
\eea

Concerning the axion production, now in addition to the misalignment mechanism discussed in
Sec.~\ref{axinorelic}, there is the production coming from the gravitino NLSP decay.
Nevertheless, the axions produced in this way will constitute
`dark radiation', i.e. ultrarelativistic and invisible species with respect to the
cold DM measured by Planck. The amount of dark radiation is under stringent constraints~\cite{Poulin:2016,Berezhiani:2015,Chudaykin:2016,Chudaykin:2017,Bringmann:2018}, and as a consequence it gives a small contribution to the DM density.

A quantity that will be useful along this work is the fraction of gravitino NLSP that decays into dark radiation.
{For that we can define}
\bea
f_{ddm}^{\text{DR}} = 
f_{3/2} \left( 1 - 
r_{\tilde a}
\right),
\label{ddmfraction}
\eea
with
\bea
f_{3/2}=\frac{\Omega_{3/2}^{\text{TP}}}{\Omega_{cdm}^{\text{Planck}}}
\label{fracgra}
\eea
the gravitino NLSP fraction.
The subscript $ddm$ denotes decaying dark matter, and DR stands for dark radiation. 
It is worth noticing here the following:
\begin{itemize}
\item Planck Collaboration obtains $\Omega_{cdm}^{\text{Planck}}h^2\simeq 0.12$ today from measurements at recombination time using the standard
$\Lambda$CDM model. We are working with decaying DM, so the cold DM density has a time dependence due to the fact that some of the gravitino NLSP energy density is `lost' as dark radiation. {Nevertheless, the latter quantity has to be small, as discussed above.}
\item Decaying DM and its fraction to dark radiation, $f_{ddm}^{\text{DR}}$, refers to the contribution of the mentioned decay of gravitino NLSP into axino LSP and axion, not to be confused with the decays of axino LSP and gravitino NLSP into photon plus neutrino.
\end{itemize}

\noindent
Let us finally point out that due to the axion-photon mixing, the axions emitted from the gravitino decay can be converted into photons in the presence of a magnetic field, potentially producing a signal. However, the conclusion of Ref.~\cite{Bae:2019} is that considering a QCD axion (as in our case), the conversion probability is too small to be observed.

\subsection{$\gamma$-ray fluxes from axino and gravitino decays}
\label{subsec: gamma-ray flux}

The constraints set to the $\gamma$-ray emission from DM considers usually that it is composed by only one particle species. In the case of DM decay, the constraints place lower limits to the particle lifetime. If axino and gravitino coexist, being one the LSP and the other the NLSP, respectively, both candidates can be sources of $\gamma$-ray radiation. Nevertheless, it is easy to normalize the signal considering that a specific source is a fraction of $\Omega_{cdm}^{\text{Planck}}h^2$.

The differential flux of $\gamma$ rays from DM decay in the Galactic halo is calculated
in Eq.~(\ref{eq:decayFlux}), with $\rho_{\text{halo}}$ as a crucial quantity.
Assuming multicomponet DM, and that the distribution of each species is homogeneous along the DM distribution, we simply have
\begin{equation}
\rho_{\text{halo}}=\sum_i \rho_{\text{DM}_i},
 \label{rhohalo}
\end{equation}
where the $i$-$th$ DM density component $\rho_{\text{DM}_i}$ can be expressed as
\begin{equation}
\rho_{\text{DM}_i}=
f_{\text{DM}_i}\ \rho_{\text{halo}},
 \label{rhoDMj}
\end{equation}
with 
\begin{equation}
f_{\text{DM}_i}\equiv  
\frac{\Omega_{\text{DM}_i}}{\Omega_{cdm}^{\text{Planck}}}.
 \label{rhoDMj2}
\end{equation}
To calculate now the $\gamma$-ray flux from the $i$-$th$ DM component 
that decays to photons, we just have to replace $\rho_{\text{halo}}\rightarrow \rho_{\text{DM}_i}$ in Eq.~(\ref{eq:decayFlux}) obtaining
the following differential flux of $\gamma$-rays:
\begin{equation}
\frac{d\Phi_{\gamma}^{\text{DM}_i}}{dEd\Omega}=f_{\text{DM}_i} \frac{d\Phi_{\gamma}^{\text{100\% DM}_i}}{dEd\Omega},
\label{fluxcase1}
\end{equation}
where $\frac{d\Phi_{\gamma}^{\text{100\% DM}_i}}{dEd\Omega}$ is the would-be differential flux if we consider $\rho_{\text{DM}_i}=\rho_{\text{halo}}$.
Finally, taking into account that the constraint to the $\gamma$-ray flux is presented as a lower limit to DM lifetime considering only one DM component, in a multicomponent scenario 
it is useful to use for each component the effective lifetime
\begin{equation}
\tau_{\text{DM}_i \text{-eff}}
= f^{-1}_{\text{DM}_i}\
\tau_{\text{DM}_i},
 \label{newtime1}
\end{equation}
where $\tau_{\text{DM}_{i} \text{-eff}}$ can be tested against the lower limit reported by the experimental collaborations.

However, we cannot apply straightforwardly the above formulas to our multicomponent DDM scenario made of axino LSP (DM$_1$) and gravitino NLSP (DM$_2$). The reason is that
their fractions
change in time due to gravitino decay into axino, so 
taking into account Eqs.~(\ref{NLSPrelicgeneral0}) and~(\ref{LSPrelicgeneral}),
we must do the following replacements in 
Eq.~(\ref{fluxcase1}) for gravitino and axino respectively:
%
%
%
\bea
f_{\text{DM}_2}  &\rightarrow & 
f_{3/2}\ 
e^{-(t_{\text{today}}-t_0)/\tau_{3/2}},
\\
\label{primera}
f_{\text{DM}_1} & \rightarrow & 
f_{\tilde a} + 
r_{\tilde a}\
f_{3/2}\ \left(1 - e^{-(t_{\text{today}}-t_0)/\tau_{3/2}}\right),
\label{fractionstime}
\eea
with $f_{3/2}$ as in Eq.~(\ref{fracgra}) and
\begin{equation}
f_{\tilde a}=  
\frac{\Omega_{\tilde a}^{\text{TP}}}{\Omega_{cdm}^{\text{Planck}}}.
 \label{rhoDMj222}
\end{equation}
As expected, if gravitino NLSP decay into axino LSP plus axion is not allowed, one gets the same result as in Eq.~(\ref{fluxcase1}). 

%
%

Finally, in a same fashion stated before, it is easier for the analysis to consider an effective lifetime in our multicomponent DDM scenario. Thus Eq.~(\ref{newtime1}) becomes
%
%
%
\bea
\tau_{3/2\text{-eff}} & = & 
\left(
f_{3/2}\ 
e^{-(t_{\text{today}}-t_0)/\tau_{3/2}}\right)^{-1} 
\Gamma^{-1}(\psi_{3/2}\rightarrow\gamma\nu_i),
 \label{newtime2}
 \\
 \tau_{\tilde a\text{-eff}} & = & 
\left[
f_{\tilde a} + 
r_{\tilde a}\
f_{3/2}\ \left(1 - e^{-(t_{\text{today}}-t_0)/\tau_{3/2}}\right)\right]^{-1} 
\Gamma^{-1}(\tilde a\rightarrow\gamma\nu_i),
 \label{newtime22}
\eea
where in the first equation the 
BR($\psi_{3/2}\rightarrow\gamma\nu_i$) $\simeq \Gamma (\psi_{3/2}\rightarrow\gamma\nu_i)/\Gamma (\psi_{3/2}\rightarrow a \tilde a)$
has been taken into account.

It is now straightforward to apply the analyses of these Subsections  to study the current constraints on the parameter space of our scenario, as well as the prospects for its detection.
For simplicity, in what follows we will use $t_0=0$ for the computation.


\begin{figure}[t!]
\begin{center}
 \begin{tabular}{cc}
 \hspace*{-4mm}
 \epsfig{file=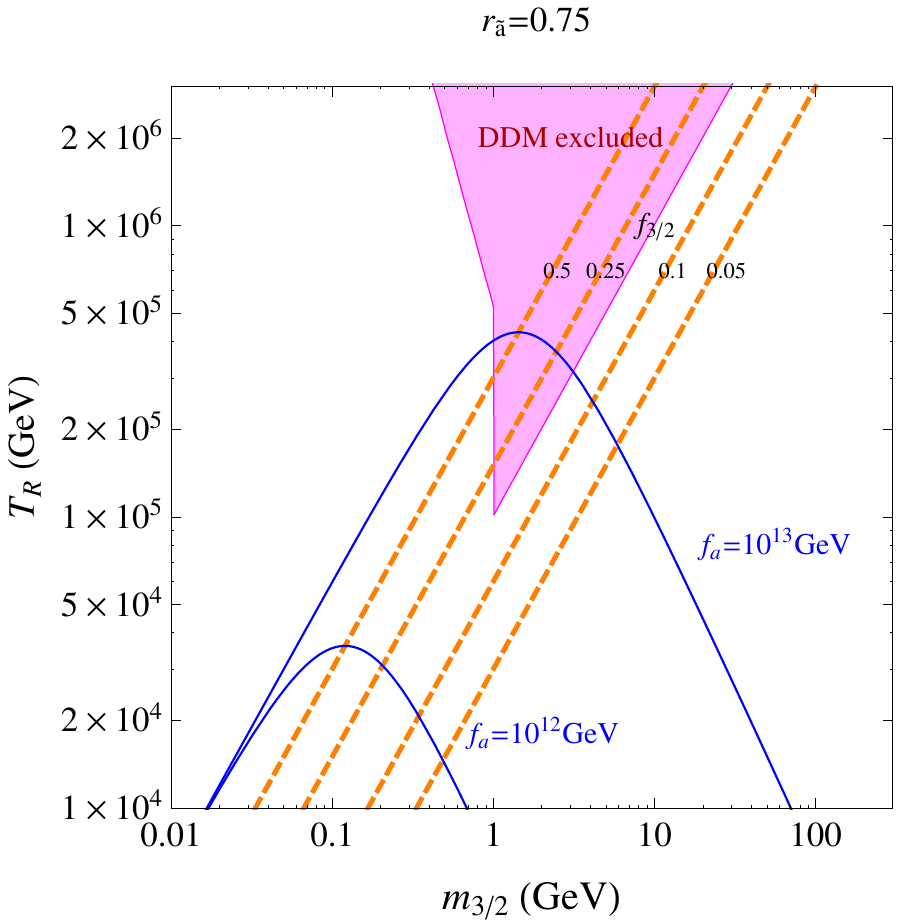,height=7cm} 
       \epsfig{file=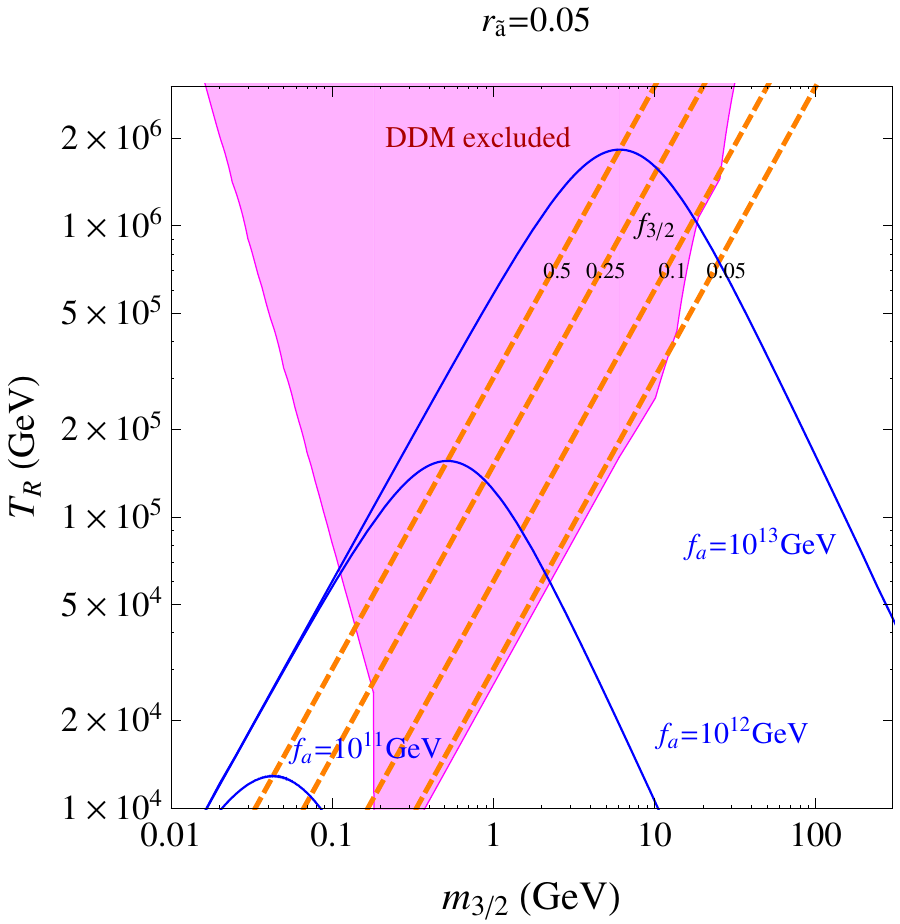,height=7cm}   
    \end{tabular}
    \captions{Constraints on the reheating temperature versus gravitino NLSP mass for the multicomponent DDM scenario with axino LSP, and mass relations
$r_{\tilde a} = 0.75, 0.05$ in left and right panels respectively. Blue lines correspond to points with $\Omega_{3/2}h^2+\Omega_{\tilde{a}}h^2$ equal to $\Omega_{cdm}^{\text{Planck}}h^2$ at recombination era in agreement with Planck observations,
for several values of the PQ scale, $f_a=10^{11}, 10^{12}, 10^{13}$ GeV. For a given $f_a$, the region above the corresponding blue line is excluded by overproduction of cold DM. 
The magenta region is excluded by cosmological observations for DDM models~\cite{Poulin:2016,Berezhiani:2015,Chudaykin:2016,Chudaykin:2017,Bringmann:2018}, considering the bound on $f_{ddm}^{\text{DR}}$. Orange dashed lines correspond to the gravitino NLSP fractions, $f_{3/2}=
0.5, 0.25, 0.1, 0.05$.
The lower bound $m_{3/2}\gsim 0.017$ GeV is obtained from Eq.~(\ref{relicgravitinos}) assuming the conservative limit $T_R \gtrsim 10^4$ GeV.
}
    \label{figaxinoLSP1}
\end{center}
\end{figure}

\subsection{Results 
}
\label{subsec: Results}

\subsubsection{Constraints from cosmological observations}

To analyze the regions of the parameter space that can satisfy the current experimental constraints on DDM models, similar as in Ref.~\cite{Hamaguchi:2017} we show $T_R$ versus $m_{3/2}$ in Fig.~\ref{figaxinoLSP1} for our DDM scenario with axino LSP and gravitino NLSP. In the left panel we use the mass relation 
$r_{\tilde a}=0.75$,
whereas in the right panel $r_{\tilde a}=0.05$.

The blue lines show points of the parameter space with $\Omega_{3/2}h^2+\Omega_{\tilde{a}}h^2$ fulfilling Planck observations at recombination era, 
for different PQ scales. 
For a given $f_a$, 
the region above the corresponding blue line is excluded by overproduction of cold DM. The region below could be allowed if we assume another DM contribution, e.g. axions from misalignment production. Although this might be an interesting scenario, a third cold DM candidate is beyond the scope of this work, so we will focus on values of the parameters fulfilling the blue contour.
On the other hand,
the orange dashed lines correspond to different values of the gravitino NLSP fraction 
$
f_{3/2}$.

Note that using Eq.~(\ref{decaytoaa}) we can define three different regions in the figure,
according to whether the decay of gravitino NLSP into axino LSP plus axion takes place after the present era, between recombination and the present era or before recombination.
For example, in the right panel there is a
small mass region or long-lived gravitino NLSP for $m_{3/2}\lesssim 0.2$ GeV, an intermediate mass region for $0.2 \lesssim m_{3/2}\lesssim 10$ GeV, and a large mass region or short-lived gravitino NLSP for $m_{3/2} \gtrsim 10$ GeV. 


Finally, the magenta regions in both panels are excluded by cosmological observations~\cite{Poulin:2016,Berezhiani:2015,Chudaykin:2016,Chudaykin:2017,Bringmann:2018}, taking into account the stringent constraints on the fraction of gravitino NLSP relic density that decays to dark radiation, $f_{ddm}^{\text{DR}}$. 
These constraints are usually presented as upper limits for this fraction.
For the intermediate gravitino NLSP mass region one obtains~\cite{Poulin:2016,Berezhiani:2015,Chudaykin:2016,Chudaykin:2017,Bringmann:2018} 
$f_{ddm}^{\text{DR}} \lesssim 0.042$,
and the corresponding ones for small and large mass regions can be found in 
Ref.~\cite{Poulin:2016}.
As can be seen in Fig.~\ref{figaxinoLSP1}, these constraints allow different values of $f_{3/2}$ depending on the relation between the axino and gravitino masses $r_{\tilde a}
$ (see Eq. (\ref{ddmfraction})). 
Recall that $f_{ddm}^{\text{DR}}$ measures the allowed energy density lost as ultrarelativistic species. For the decay of gravitino NLSP to axino LSP plus axion, the latter will always be ultrarelativistic but the behaviour of the non-thermally produced axino depends on 
$r_{\tilde a}$.

\begin{figure}[t!]
\begin{center}
 \begin{tabular}{cc}
 \hspace*{-4mm}
 \epsfig{file=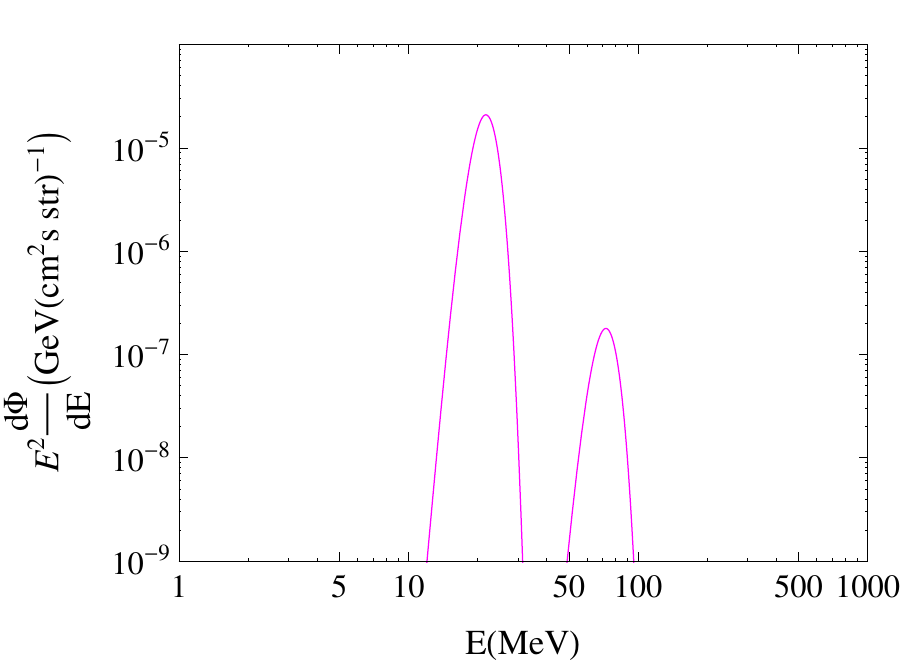,height=7cm} 
    \end{tabular}
    \captions{Double $\gamma$-ray line generated by the decay of a coexisting axino-gravitino mixture as the DM of the Universe. The spectral features are created by an axino LSP of 43.5 MeV and a gravitino NLSP of 145 MeV decaying into a photon-neutrino pair for $f_a=10^{13}$ GeV and $\left|U_{\tilde{\gamma} \nu}\right|=10^{-6}$. 
The lines are convolved with Gaussians assuming 10\% energy resolution of the instrument. The double line plotted corresponds to the emission from $10^\circ \times 10^\circ$ square around the Galactic center. 
To set limits we use standard line search results, applying them to each line separately.
}
    \label{flux_lines}
\end{center}
\end{figure}

\begin{figure}[t!]
\begin{center}
 \begin{tabular}{cc}
 \hspace*{-4mm}
 \epsfig{file=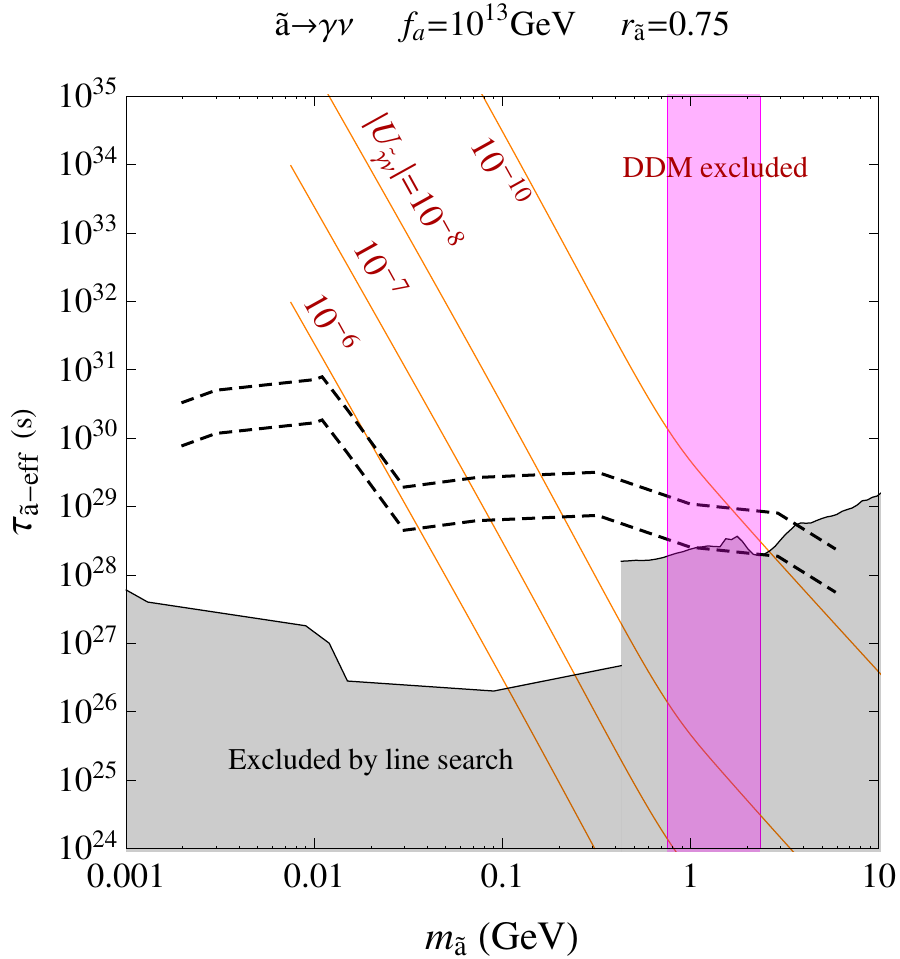,height=7cm} 
       \epsfig{file=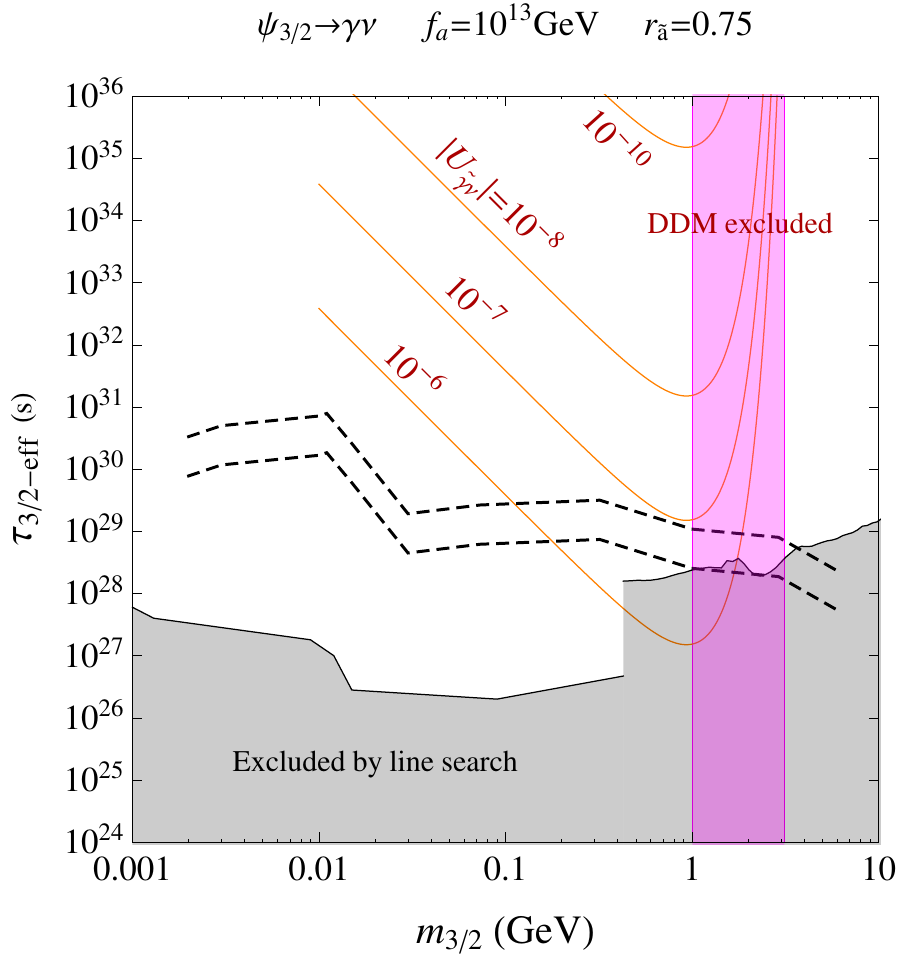,height=7cm}
       \vspace*{-0.8cm}       
   \\ & 
    \end{tabular}
    \captions{
Constraints on effective lifetime versus axino LSP mass (left panel) and gravitino NLSP mass (right panel). The $\gamma$-ray signals from axino and gravitino decays are analyzed separately in left and right panels, respectively.
The grey region below the black solid line on the left (right) is excluded by line searches in the Galactic halo by COMPTEL (\Fermi LAT~\cite{Ackermann:2015lka}). The region below the upper (lower) black dashed line could be probed by e-ASTROGAM~\cite{eAstrogamAngelis:2017} with observations of the Galactic center assuming Einasto B (Burkert) DM profile.
The
orange solid lines correspond to the predictions of the $\mn$ for several representative values of 
$|U_{\tilde{\gamma} \nu}|$, for the case $f_a=10^{13}$ GeV and $r_{\tilde a}=0.75$.
The lower bound $m_{3/2}\gsim 0.017$ GeV is obtained from Eq.~(\ref{relicgravitinos}) assuming the conservative limit $T_R \gtrsim 10^4$ GeV.
The magenta region is excluded by cosmological observations for DDM models~\cite{Poulin:2016,Berezhiani:2015,Chudaykin:2016,Chudaykin:2017,Bringmann:2018}, considering the bound on $f_{ddm}^{\text{DR}}$. 
}
    \label{figaxinoLSP2}
\end{center}
\end{figure}

\subsubsection{Constraints from $\gamma$-ray observations and prospects for detection}
\label{subsec: Prospects}

To analyze the effect of considering our multicomponent DDM scenario on the $\gamma$-ray signal, we plot in Fig.~\ref{flux_lines} an example of the spectral features generated by coexisting gravitino-axino particles.
We use the lines separately to set constraints on the model, considering that the resulting flux from each line is not affected by the other signal. As we will see below, in the case of a detectable double line, the two signals turn out to be located at different enough energies to do not overlap and give rise to two resolvable lines. For that reason, we show two panels in Fig.~\ref{figaxinoLSP2}, one for each DM particle.
The left (right) panel shows the limits on the parameter space considering the line produced by axino LSP (gravitino NLSP) decaying into $\gamma \nu$. 
In this example we fix $f_a=10^{13}$ GeV and $r_{\tilde{a}}=0.75$, which are the same values as those used for the upper blue line in the left panel of Fig.~\ref{figaxinoLSP1}. 
Thus the left and right panels of Fig.~\ref{figaxinoLSP2} correspond to the same axino LSP plus gravitino NLSP scenario, and the constraints obtained from both panels have to be taken into account for each point in the parameter space. For example, the point with $m_{3/2}=0.5$ GeV and $|U_{\tilde{\gamma} \nu}|=10^{-7}$ 
seems not to be excluded in the right panel by line searches, however it corresponds to an axino mass $m_{\tilde{a}}=r_{\tilde a}\times 0.5=0.375$ GeV which for
$|U_{\tilde{\gamma} \nu}|=10^{-7}$
is clearly excluded in the left panel.

One can see the effect that the decay of gravitino NLSP into axino LSP plus axion has in the effective lifetime of the two DM particles, comparing the left panel of Fig.~\ref{figaxinoLSP2} with the orange region 
of Fig.~\ref{figaxinoalone1}, where only the axino LSP is the DM.
For $m_{\tilde{a}}\leq 0.8$ GeV,
the effective lifetime is larger than the lifetime without gravitino NLSP, mainly due to the low axino fraction $f_{\tilde a}$ contributing to the first term of 
Eq.~(\ref{newtime22}). A similar situation occurs for $0.8\leq m_{\tilde{a}}\leq 1.2$ GeV, where the contribution to the second term of Eq.~(\ref{newtime2}) is significant due to the axino energy density from the gravitino decay. On the other hand, for $m_{\tilde{a}} \geq$ $1.2$ GeV, the gravitino decay takes place in a sufficiently early time and/or the gravitino fraction $f_{3/2}$ is low enough, in such a way that the effective lifetime is similar to the scenario of Fig.~\ref{figaxinoalone1} with initially 100\% axino DM.
The right panel of Fig.~\ref{figaxinoLSP2} shows the same parameter space but for  the effective lifetime of the gravitino NLSP. We can see the effect of the reduction of the gravitino relic density due to its decay into axino LSP for $1\leq m_{3/2}\leq 3$ GeV, as can be deduced from Eq.~(\ref{newtime2}). Lower gravitino masses imply a longer decay time into axino LSP, so that in that region we can have at the present era a DM distribution with both candidates producing a double line.

\begin{figure}[H]
 \begin{center}
  \begin{tabular}{cc}
 \hspace*{-0mm}
  	\includegraphics[height=6cm]{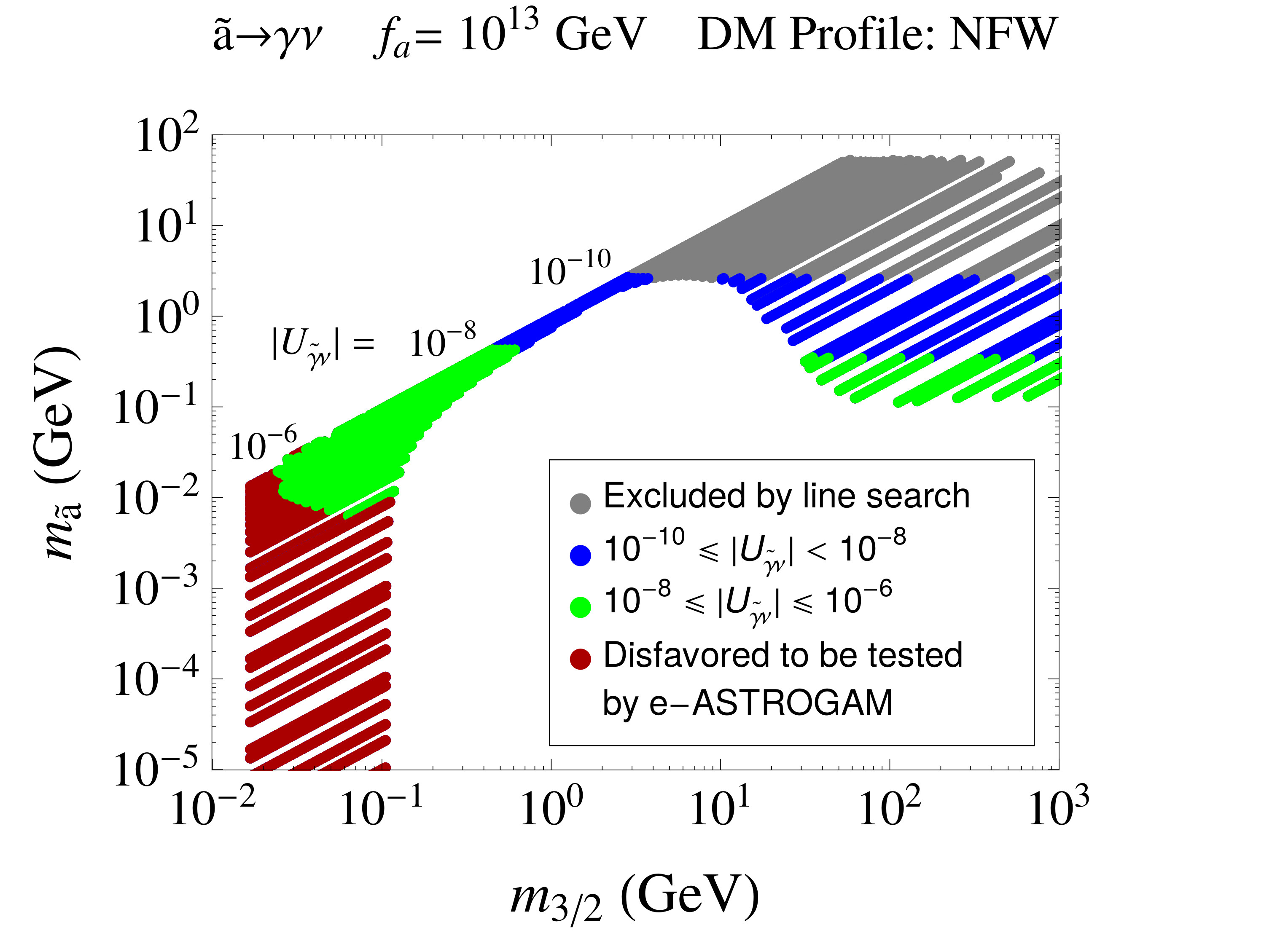} 
\hspace*{-0.5cm} \includegraphics[height=6cm]{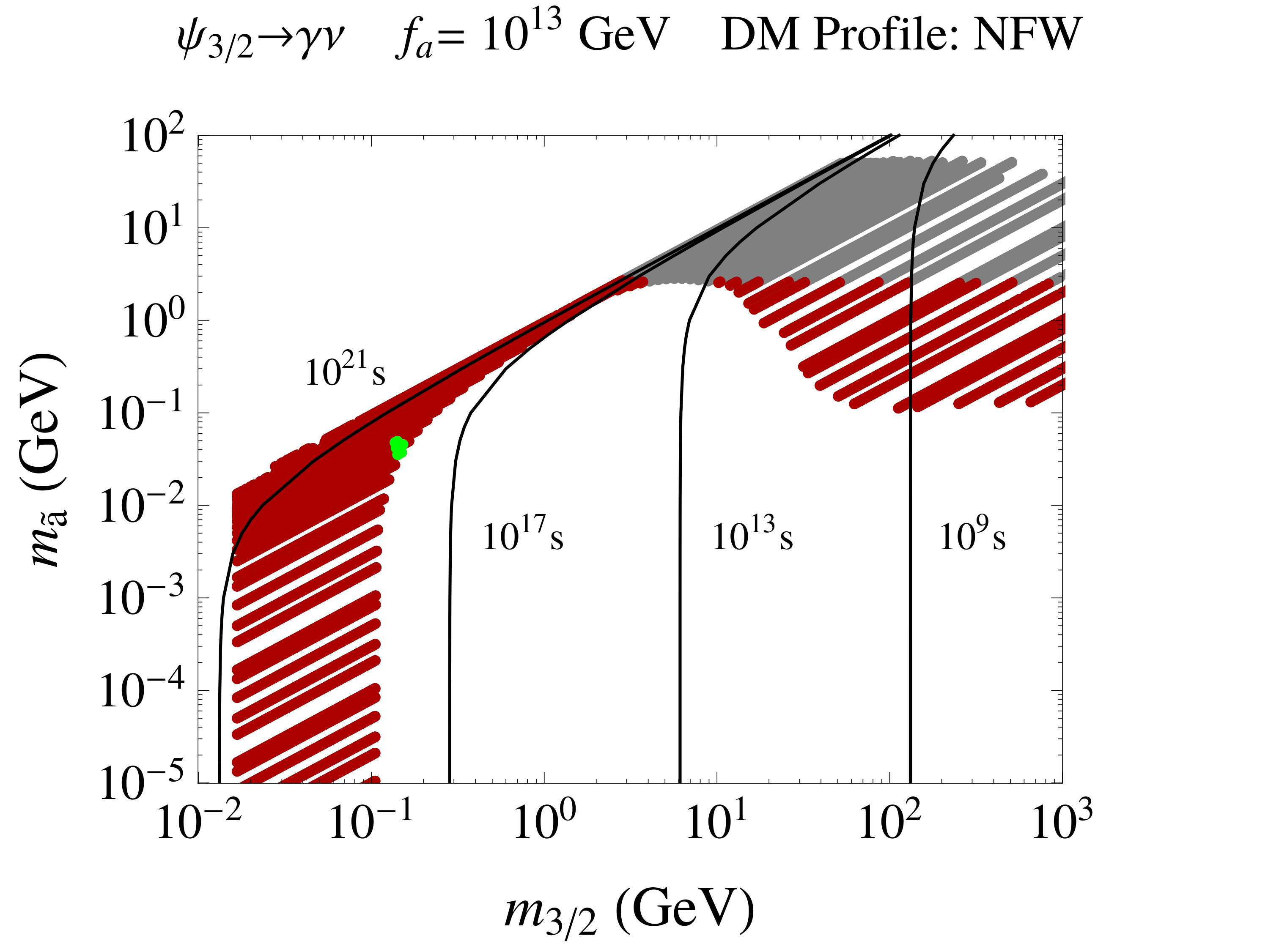}
   \\ \hspace*{-0mm} \includegraphics[height=6cm]{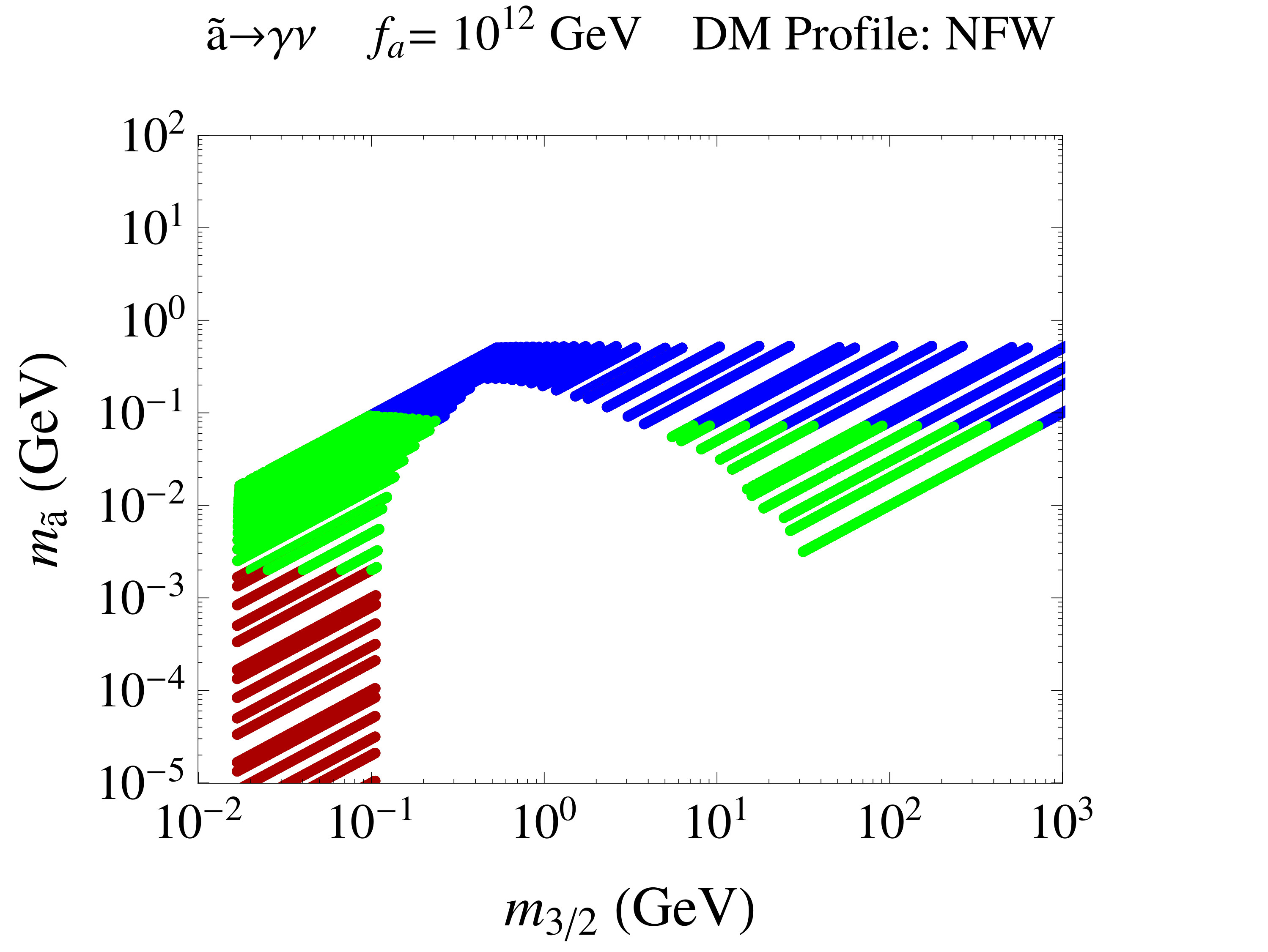} \hspace*{-0.5cm} \includegraphics[height=6cm]{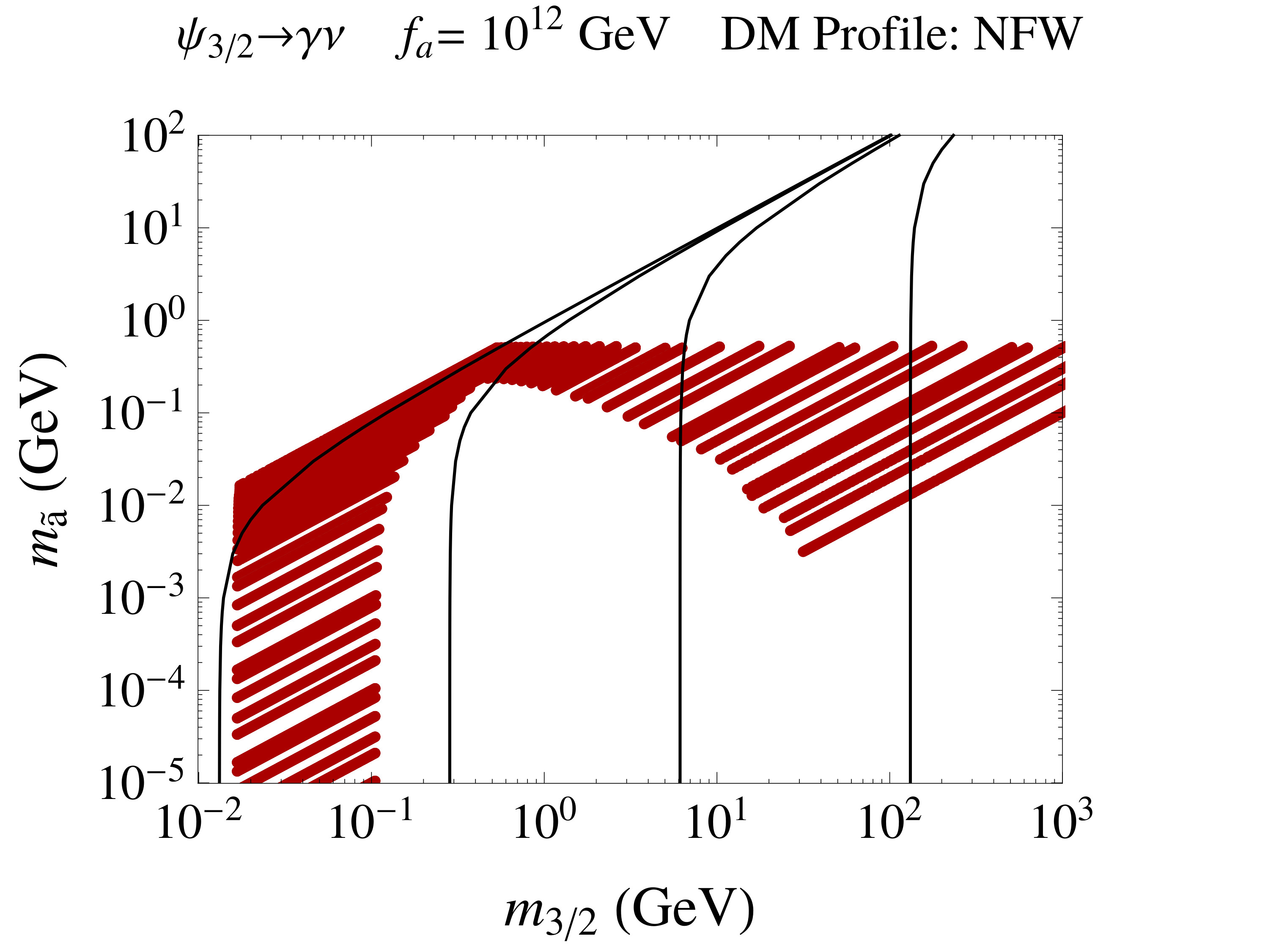}\\     
	\hspace*{-0mm}  \includegraphics[height=6cm]{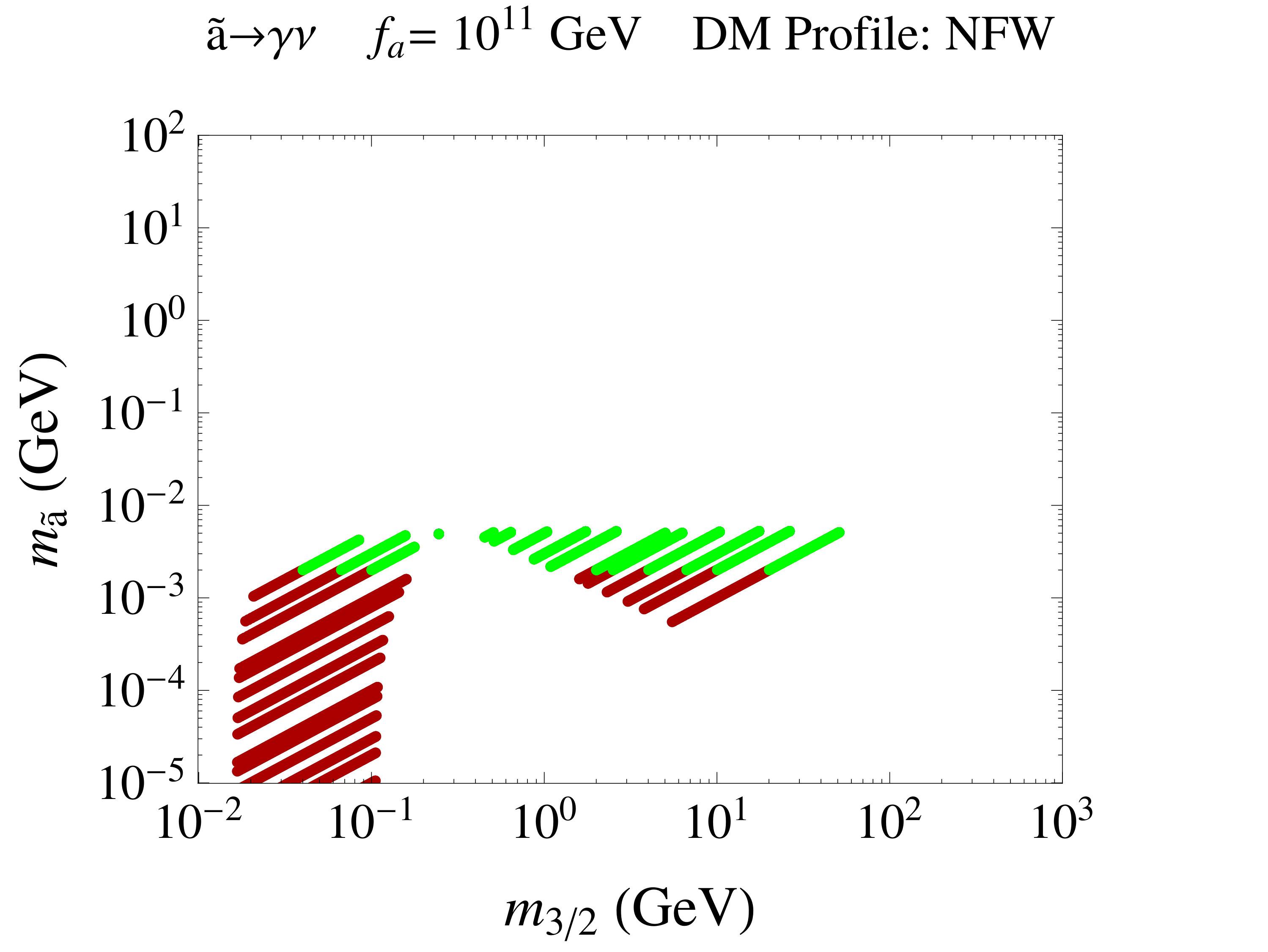} \hspace*{-0.5cm}
       \includegraphics[height=6cm]{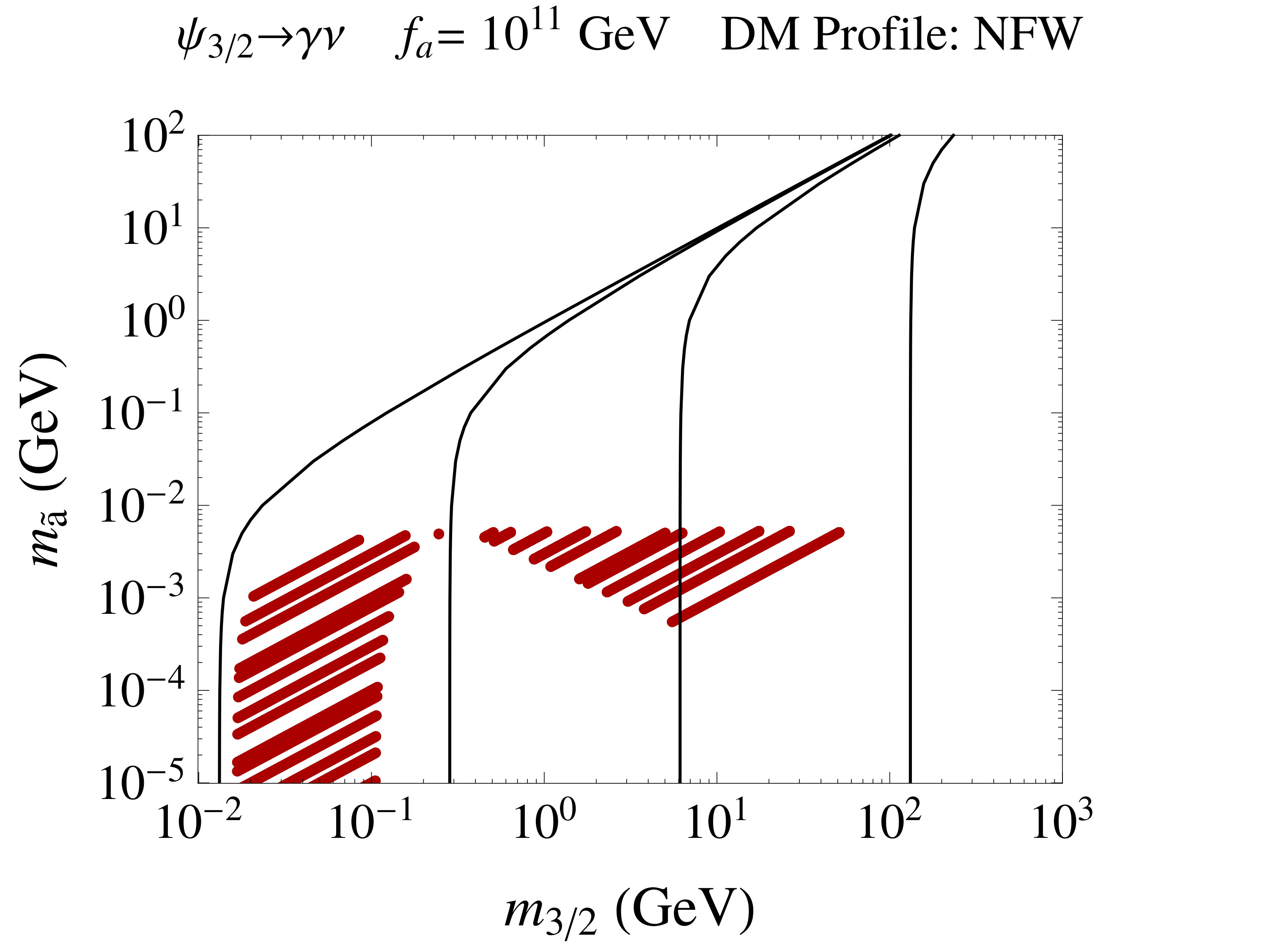} 
  \end{tabular}
  \captions{\small 
Constraints on axino LSP mass versus gravitino NLSP mass in the range
$10^{-4} \leq r_{\tilde a} \leq 0.95$. The
$\gamma$-ray signals from 
axino 
and 
gravitino 
decays are analyzed separately in left and right panels, respectively, assuming a NFW profile.
The grey region corresponds to points excluded by line searches in the Galactic halo by COMPTEL and \Fermi LAT~\cite{Ackermann:2015lka}.
Blue and green regions correspond to points that could be probed by e-ASTROGAM for 
two representative ranges of $|U_{\tilde{\gamma}\nu}|$ in the $\mn$.
In the top left panel, the values in the border between regions are labeled, and for the rest of the panels the labeling is the same.  
If the same point can be probed in both panels, a double-line signal could be measured.
The red region corresponds to points disfavored to be tested by e-ASTROGAM.
In the right panels, 
the black solid lines show different values of 
$\tau_{3/2}\simeq {\Gamma}^{-1} (\psi_{3/2} \rightarrow a \, \tilde{a})$.
All the points shown satisfy 
$\Omega_{3/2}h^2+\Omega_{\tilde{a}}h^2$ equal to $\Omega_{cdm}^{\text{Planck}}h^2$ at recombination era in agreement with Planck observations, as well as DDM constraints for  $f_{ddm}^{\text{DR}}$.
}
    \label{figconstrains1}
\end{center}
\end{figure}

To carry out a complete analysis of the allowed parameter space, we have performed a scan over the following range:
\begin{equation}
10^{-4} \leq r_{\tilde a} \leq 0.95.
\end{equation}
The result is shown in
Fig.~\ref{figconstrains1}, where the
$\gamma$-ray signals from axino and gravitino decays are analyzed separately in left and right panels, respectively.
Green and blue regions correspond to points that could be probed with the projected sensitivity of e-ASTROGAM assuming a NFW profile and a region of interest of 10$^\text{o} \times$10$^\text{o}$ around the Galactic center, for different values of the photino-neutrino mixing parameter $|U_{\tilde{\gamma} \nu}|$.
In particular, the green points
correspond to the most natural range for $|U_{\tilde{\gamma} \nu}|$, as discussed in Eq.~(\ref{relaxing}). It is worth mentioning here that this range includes the typical parameter space that can reproduce the observed neutrino physics in bilinear RPV models, thus the constraints obtained also apply to those models.

As we can see in the figure,
for values of $r_{\tilde a}$ close to 1, i.e. the narrow allowed region, we recover the allowed parameter space obtained for axino LSP as the only DM, i.e. without the gravitino NLSP effect. This is because the DDM constraints for $f_{ddm}^{\text{DR}}$ become relaxed since
${\Gamma} (\psi_{3/2} \rightarrow a \, \tilde{a})  \rightarrow 0$ when
$r_{\tilde a} \rightarrow 1$ (see Eq.~(\ref{decaytoaa})).
The remaining effects concerning the $\gamma$-ray fluxes are just given
by the relic density fractions of the LSP and NSLP.
For lower values of $r_{\tilde a}$,
the allowed parameter space is modified due to the DDM constraints, giving rise to the two separated allowed mass regions 
shown 
in Fig.~\ref{figconstrains1}.
In this sense, note that the DDM exclusion in Fig.~\ref{figaxinoLSP1} for 
$r_{\tilde a}=0.75$ and $f_a=10^{13}$~GeV, leaves two allowed branches for the blue solid line ($m_{3/2} \lsim 1$ GeV and $m_{3/2} \gtrsim 3$ GeV) with the correct relic density.

{From Fig.~\ref{figconstrains1}, we can conclude that a significant region of the parameter space of our DDM scenario, inside the mass ranges $7 \, \text{MeV} \lsim m_{\tilde a} \lsim 3$ GeV and $20 \, \text{MeV} \lsim m_{3/2} \lsim 1$ TeV,
could be tested by next generation $\gamma$-ray telescopes, and this is specially true thanks to the line signal coming from axino LSP (left panels).
Note in this sense that axino and gravitino decay widths into photon-neutrino which are relevant quantities for the amount of photon flux 
(see Eqs.~(\ref{eq:decayFlux}),~(\ref{newtime2}) and~(\ref{newtime22})) 
fulfill $\Gamma(\tilde{a}\rightarrow\gamma\nu_i) > \Gamma(\psi_{3/2}\rightarrow\gamma\nu_i)$ within these mass ranges for the values of $r_{\tilde a}$ discussed in Sect.~\ref{subsec:Gravitino decay}.
For smaller values of the mass ratio $r_{\tilde a}$, 
one would expect an important flux attributed to gravitino decay.
However, due to the dark-radiation exclusion region the mass of gravitinos in this case has to be very large implying that they already decayed into axion-axino before $t_{\text{today}}$ (see in the right panels that these are points to the right of the black line corresponding to $10^{17}$ s),
or very small $m_{3/2} \lsim 0.1$ GeV, implying that the photon flux is small 
(see Eq.~(\ref{gravitinolifetime_LSP})).
According to this, we also expect a line signal coming from gravitino NLSP to be measured in a smaller region. This is actually the green region of the top right panel
corresponding to $f_a=10^{13}$~GeV, with masses $m_{3/2}\sim 150$ MeV and $m_{\tilde a}\sim 40$ MeV.
Since the same points can be probed in both, left and right panels, a double-line signal could be measured as a overwhelming smoking gun.
It is worth pointing out that the black solid lines in the right panels show us, as expected,  
that this detectable $\gamma$-ray signal from gravitino NLSP decay
lies in the region of the parameter space with $\tau_{3/2}\simeq {\Gamma}^{-1} (\psi_{3/2} \rightarrow a \, \tilde{a})~>~t_{\text{today}}$.}


Note that within this scenario, there is an important region of the detectable parameter space where heavy gravitino masses are allowed, $m_{3/2}> 10$ GeV.
In this region, besides the studied photon-neutrino channel, other decay modes become relevant as those 
involving $Z$, $W$ and Higgs bosons in two and three-body decays~\cite{GomezVargas:2017}. This results in an increase of the gravitino decay width to visible particles, producing the injection of energetic hadronic and electromagnetic species in the early Universe, that could alter the big bang nucleosynthesis (BBN) process  or the cosmic microwave background (CMB) spectrum. 
Nevertheless, in the gravitino mass region where this occurs, the dominant decay process is gravitino NLSP to axino LSP plus axion,
as can be seen comparing the right panels of Figs.~\ref{figconstrains1} and~\ref{figaxinoLSP2}.
Another important factor to take into account is that for the heavy gravitino mass region, its thermal relic density is low 
(see the orange dashed lines in Fig.~\ref{figaxinoLSP1}). Therefore, there is no significant energy deposited to the visible sector during 
the early Universe due to the gravitino energy density reduction, and for the 
$\gamma$-ray analysis it is safe to consider only the gravitino to photon-neutrino process.


\begin{figure}[t!]
 \begin{center}
  \begin{tabular}{cc}
 \hspace*{-4mm}
 \includegraphics[height=6cm]{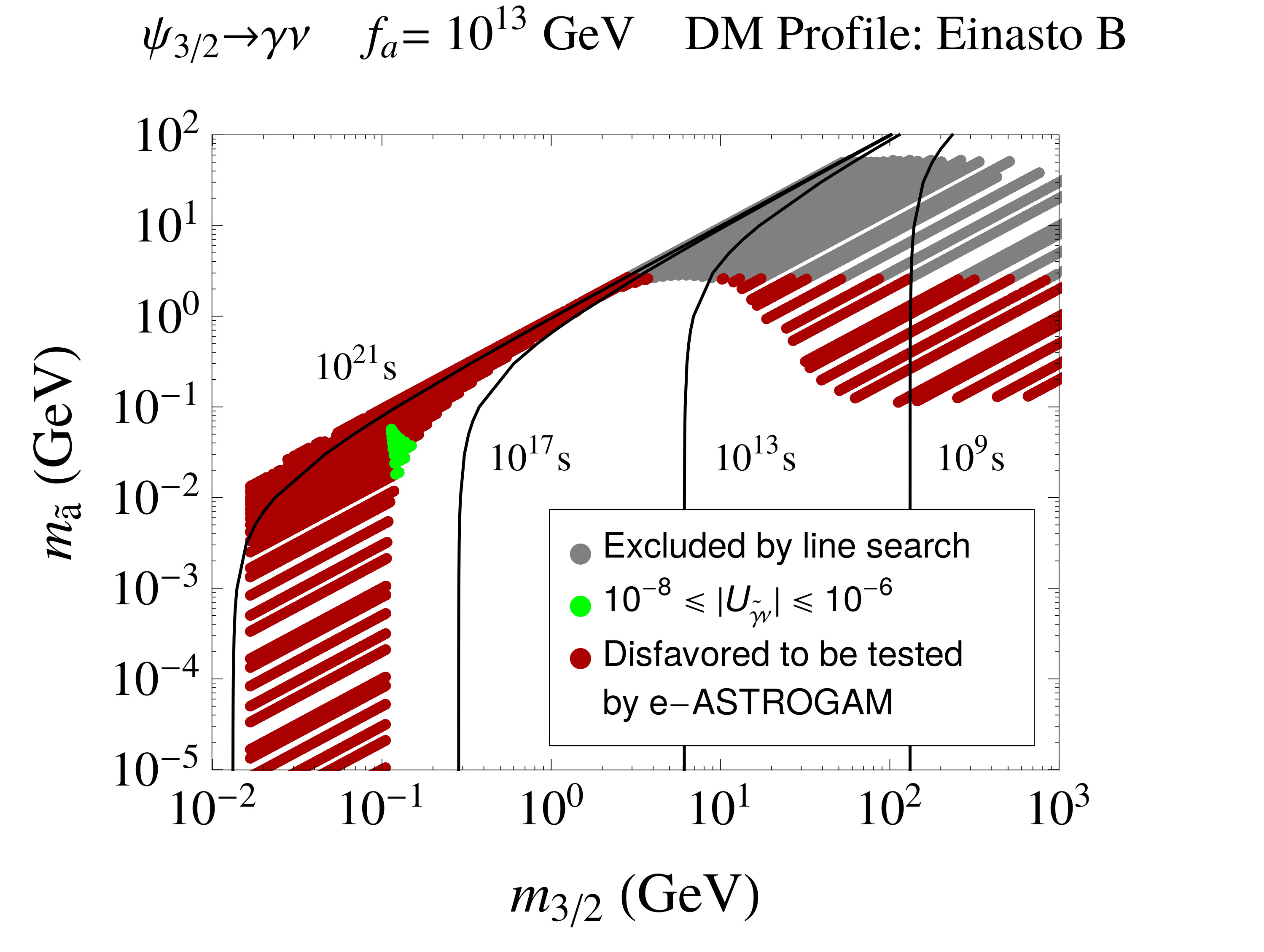} 
  \end{tabular}
  \captions{The same as in Fig.~\ref{figconstrains1}, but showing only the $\gamma$-ray signal from gravitino decay into photon-neutrino and assuming an Einasto B profile.
}
    \label{figprofiles11}
\end{center}
\end{figure}

%

On the other hand,
{as already mentioned, in Fig.~\ref{figconstrains1} we used the projected e-ASTROGAM sensitivity assuming a NFW profile.
In Fig.~\ref{figprofiles11}, we show for a different DM profile, Einasto B, the detectable parameter space considering a signal coming from gravitino NLSP decay into photon plus neutrino, and a PQ scale $f_a=10^{13}$~GeV. 
Comparing this figure with the top right panel of Fig.~\ref{figconstrains1}, we can see 
that the green region to be probed through a double-line signal is slightly extended inside the ranges 
$100\lsim m_{3/2}\lsim 200$ MeV and $10 \lsim m_{\tilde a}\lsim 60$ MeV.
We have checked that the other panels in Fig.~\ref{figconstrains1} are basically not modified by this astrophysical uncertainty.
}

Let us finally point out that our analysis is based on the sensitivity of planned experiments. Nevertheless, taking into account the astonishing advances in techniques and technology of recent years, the situation could be in the future even better with respect to the one described here, potentially leading to an increase of the detectable regions, such as e.g. the green region to be probed through a double-line signal. In this sense, we have adopted here a conservative  viewpoint, leading to the presented figures and results.

\section{Conclusions}
\label{sec:conclusions}

In this work we have assumed first that the axino is the LSP and the only DM particle in the framework of the $\mn$. We have discussed its decay rate into photon plus neutrino, which is suppressed by the large PQ scale
$10^{11} \leq f_a \leq 10^{13}$ GeV and the small RPV mixing parameter 
$10^{-10} \lesssim |U_{\widetilde{\gamma}\nu}| \lesssim 10^{-6}$,
giving rise to an axino lifetime longer than the age of the Universe. For the latter result, the small values of neutrino Yukawas in the generalized electroweak-scale seesaw of the $\mn$ are crucial, determining the small values of
$|U_{\widetilde{\gamma}\nu}|$.

The corresponding relic density has also been discussed, and assuming a conservative lower bound on the reheating temperature of
$T_R \gtrsim 10^4$ GeV
an upper bound on the axino mass of $m_{\tilde a} \lsim 50$ GeV was obtained.

Then
we have studied the $\gamma$-ray flux produced in this scenario,
finding that masses
$m_{\tilde a} \gsim 3$ GeV are already excluded by {\it Fermi}-LAT searches of lines in
the Galactic halo.
Proposed MeV-GeV missions such as e-ASTROGAM would allow to explore the ranges
$2$ MeV $\lsim m_{\tilde a} \lsim 3$ GeV, 
$2\times 10^{26}\lsim \tau_{\tilde a} \lsim 8\times 10^{30}$ s, from searches in 
a ROI
around the Galactic center.
 
Second, we have analyzed the possibility of a gravitino NLSP having a large RPC partial decay width into axino LSP plus axion, in addition to the small RPV partial decay width into photon plus neutrino.
We have discussed three relevant regions, a small mass region or long-lived gravitino NLSP decaying after the present era, and intermediate mass region with the gravitino decaying between recombination and the present era, and a large mass region or short-lived gravitino NLSP decaying before recombination. 
If axino and gravitino coexist, both DM particles can be sources of $\gamma$-ray radiation.

Assuming also in this scenario $T_R \gtrsim 10^{4}$ GeV, a lower bound on the gravitino mass of $m_{3/2} \gsim 0.017$ GeV is obtained. We have also found the regions of the parameter space excluded by cosmological observations, considering the stringent constraints on the fraction of gravitino NLSP relic density that decays to dark radiation (see Fig.~\ref{figaxinoLSP1}).

Concerning the $\gamma$-ray flux produced in this DDM scenario of the $\mn$, significant regions of the parameter space could be tested by e-ASTROGAM inside the mass ranges 
$7 \: \text{MeV}  \lsim m_{\tilde a} \lsim 3$ GeV and $20 \: \text{MeV} \lsim m_{3/2} \lsim 1$ TeV. 
This is specially true thanks to the line signal coming from axino 
LSP decay (see left panels of Fig.~\ref{figconstrains1}). 
For $\tau_{3/2}\simeq {\Gamma}^{-1} (\psi_{3/2} \rightarrow a \, \tilde{a})~>~t_{today}$, a signal coming from gravitino NLSP could be measured for a narrow region of the parameter space with $f_a=10^{13}$~GeV inside the mass ranges $100\lsim m_{3/2}\lsim 200$ MeV and $10 \lsim m_{\tilde a}\lsim 60$ MeV (see the top right panel of Fig.~\ref{figconstrains1} and Fig.~\ref{figprofiles11} for NFW and Einasto B DM profile, respectively). In this case a double-line signal from axino and gravitino decays could be measured as a overwhelming smoking gun.



\section*{Acknowledgments}

The work of GAGV was supported by Programa FONDECYT Postdoctorado under grant 3160153. The work of DL and AP was supported by the Argentinian CONICET, and they also acknowledge the support through PIP 11220170100154CO. 
The work of CM was supported in part by the Spanish Agencia Estatal de Investigaci\'on 
through the grants FPA2015-65929-P (MINECO/FEDER, UE), PGC2018-095161-B-I00 and IFT Centro de Excelencia Severo Ochoa SEV-2016-0597.
We also acknowledge the support of the Spanish Red Consolider MultiDark FPA2017-90566-REDC. 
CM and DL gratefully acknowledge the hospitality of the Institut Pascal during the Paris-Saclay Astroparticle Symposium 2019, supported by P2IO (ANR-11-IDEX-0003-01 and ANR-10-LABX-0038), in whose stay the last stages of this work were carried out.

\bibliographystyle{utphys}
\bibliography{munussm}

\end{document}